\documentclass[aps,superscriptaddress,twocolumn,a4paper]{revtex4}
\usepackage[colorlinks=true, pdfstartview=FitV, linkcolor=blue, citecolor=red, urlcolor=black]{hyperref}
\usepackage{graphicx}
\usepackage[all]{xy}
\usepackage{amsmath}
\usepackage{amssymb}
\usepackage{tensor}
\newcommand{\be}{\begin{equation}}
\newcommand{\ee}{\end{equation}}
\newcommand{\ben}{\begin{eqnarray}}
\newcommand{\een}{\end{eqnarray}}
\newcommand{\bes}{\begin{subequations}}
\newcommand{\ees}{\end{subequations}}
\def\bal#1\eal{\begin{align}#1\end{align}}
\newcommand{\ov}{\overline}
\newcommand{\wt}{\widetilde}
\newcommand{\nn}{\nonumber\\}
\newcommand{\bfi}{\begin{figure}}
\newcommand{\efi}{\end{figure}}
\newcommand{\bc}{\begin{center}}
\newcommand{\ec}{\end{center}}

\newcommand{\LL}{{\cal L}}

\newcommand{\Dc}{{\cal D}}

\newcommand{\p}{\partial}

\newcommand{\vphi}{\varphi}
\newcommand{\vphic}{\ov{\varphi}}
\newcommand{\vphia}{|\varphi|}

\begin{document}

\title{Vortices in Maxwell-Chern-Simons-Higgs models with nonminimal coupling}

\author{I. Andrade}\affiliation{Departamento de F\'\i sica, Universidade Federal da Para\'\i ba, 58051-970 Jo\~ao Pessoa, PB, Brazil}
\author{D. Bazeia}\affiliation{Departamento de F\'\i sica, Universidade Federal da Para\'\i ba, 58051-970 Jo\~ao Pessoa, PB, Brazil}
\author{M.A. Marques}\affiliation{Departamento de F\'\i sica, Universidade Federal da Para\'\i ba, 58051-970 Jo\~ao Pessoa, PB, Brazil}
\author{R. Menezes}\affiliation{Departamento de Ci\^encias Exatas, Universidade Federal
da Para\'{\i}ba, 58297-000 Rio Tinto, PB, Brazil}
\affiliation{Departamento de F\'\i sica, Universidade Federal da Para\'\i ba, 58051-970 Jo\~ao Pessoa, PB, Brazil}

\begin{abstract}
We investigate the presence of vortex configurations in generalized Maxwell-Chern-Simons models with nonminimal coupling, in which we introduce a function that modifies the dynamical term of the scalar field in the Lagrangian. We first follow a route already considered in previous works to develop the Bogomol'nyi procedure, and, in this context, we use the first order equations to obtain a vortex with a novel behavior at its core. We then go further and introduce a novel procedure to develop the Bogomol'nyi methodology. It supports distinct first order equations, and we then investigate another model, in which the vortex may engender inversion of the magnetic flux, an effect with no precedents in the study of vortices within the nonminimal context.
\end{abstract} 

\vspace{1cm}

\maketitle
\section{Introduction}
Vortices are defect structures that appear in high energy physics in $(2,1)$ flat spacetime dimensions. The first relativistic model that support such configurations was suggested by Nielsen and Olesen in Ref.~\cite{NO} in 1973, with the action of a complex scalar field minimally coupled to a gauge field under a $U(1)$ symmetry, with the standard covariant derivative. An interesting feature of these structures is the absence of electric charge and the quantised character of the magnetic flux. The equations of motion that describe vortex configurations are of second order and by minimising the energy of the system, Bogomol'nyi found first order equations compatible with them in Ref.~\cite{bogopaper}.

The model proposed in Ref. \cite{NO} has the dynamics of the gauge field controlled by a Maxwell term. Nevertheless, one can exchange it for the Chern-Simons term, as suggested in Refs.~\cite{jackiw1,jackiw2,coreanos}. The vortex configurations with topological nature in this model are electrically charged, such that the electric charge and the magnetic flux are both quantised. Vortices in models with both Maxwell and Chern-Simons terms were considered in Ref.~\cite{paulkhare}. In this case, considering the scenario in which the fields are minimally coupled, in order to develop the Bogomol'nyi procedure, one must add a neutral field \cite{nmcs,bazeiamcs}. Even so, one cannot obtain a set of first order equations that completely describes the problem.

Since we are working with planar systems, we can add an anomalous magnetic moment contribution to the covariant devivative, making the coupling between the gauge and scalar fields nonminimal. The point is that the dual of $F^{\mu\nu}$ is a vector in $(2,1)$ spacetime dimensions, that is, $F_\mu=(1/2)\epsilon_{\mu\nu\lambda}F^{\nu\lambda}$, where $F_{\mu\nu}= \partial_\mu A_\nu-\partial_\nu A_\mu$, with $A_\mu$ standing for the gauge field; thus, we can change the covariant derivate from its minimal coupling form $D_\mu=\partial_\mu+ ie A_\mu$ the the nonminimal coupling described by ${\cal D}_\mu=\partial_\mu+ ie A_\mu-i q F_\mu$ \cite{Soro,AA,AB}. This possibility was considered before in Refs. \cite{torres,ghoshplb,ghosh} as a way to circumvent the presence of the additional neutral field that appeared in \cite{nmcs,bazeiamcs}: in Ref. \cite{torres} the author considered a nonminimal coupling, with the inclusion of an anomalous magnetic contribution. By doing so, he was able to obtain a set of first order equations that completely solve the equations of motion. However, the solutions engendered the nontopological character. For this reason, in Refs.~\cite{ghoshplb,ghosh} another line of investigation was considered, with the addition of a generalized magnetic permeability and a function to control the anomalous magnetic contribution, both depending only on the scalar field. When these functions are constrained in a specific manner, it is possible to develop the Bogomol'nyi procedure and obtain first order equations. In this model, the form of the aforementioned functions may lead to nontopological and/or topological configurations whose charge is proportional to the magnetic flux. The physical properties of planar systems have a long history, and interesting lines of investigations concerning fractional statistics and anyons appeared before, for instance, in \cite{Wil,Kha} and in references therein.    

The presence of nonminimal coupling may be used to get the Chern-Simons term by spontaneous symmetry breaking in a Maxwell-Higgs model \cite{nm0}. Over the years, in the context of models with nonminimal coupling between the scalar and the gauge field, several works appeared in the literature; see Refs.~\cite{nm1,nm2,nm3,nm4,nm5,nm6,nm7,nm8,nm9,nm10,nm11,nm12}. In particular, in Ref.~\cite{nm3}, vortex configurations were investigated in model with non-Abelian fields. Their associated magnetic flux is not quantised due to their nontopological nature. However, both the electric charge and angular momentum are quantised. In Refs.~\cite{nm4,nm8,nm10}, nonrelativistic models were studied, and in Refs.~\cite{nm7,nm11,nm12}, the authors investigated vortices in $O(3)$-sigma models, which may support both topological and non-topological profile. 

In this paper, we investigate a generalized model, with the dynamical term of the scalar field containing a function of the scalar field in the nonminimal coupling. This is explained in the next section, where we calculate some properties of the model, such as the equations of motion, the current and the energy momentum tensor. We then focus on developing a first order formalism to describe the vortex configurations of interest in Sec. \ref{secthree}. In Sec.~\ref{firstcase}, we follow a path similar to the one suggested in Refs.~\cite{ghoshplb,ghosh} and develop the Bogomol'nyi procedure for this case by minimising the energy of the system. We provide an example to illustrate how the aforementioned function that drives the dynamical term of the scalar field plays a role in the profile of the solutions. In Sec.~\ref{secondcase}, we introduce a novel procedure to get a first order formalism for the model. We provide two examples that present novel physical features in the considered scenario, such as the absence of the monotonic behavior of the solutions and magnetic flux inversion, an effect that appeared before in other contexts, in particular in the case of fractional vortices in two-component superconductors \cite{fluxprl}, and also in models with breaking of the Lorentz invariance \cite{casanaflux}. We conclude the investigation in Sec. \ref{end}, where we comment on the main results obtained in the work and on several possibilities of investigations related to the presence of the generalized nonminimal coupling considered in the present study.  

\section{The model}
We consider a gauge field and a complex scalar field in $(2,1)$ flat
spacetime dimensions, with metric $\eta_{\mu\nu}=\text{diag}(+,-,-)$ and action $S=\int d^3x\,\LL$, where the Lagrange density is
\be\label{lagrange}
\begin{aligned}
\LL =& -\frac14P(\vphia)F_{\mu\nu}F^{\mu\nu} + \frac{\kappa}{4}\epsilon^{\lambda\mu\nu}A_\lambda F_{\mu\nu}\\ 
&+M(\vphia)\ov{\Dc_\mu\vphi}\Dc^\mu\vphi - V(\vphia).
\end{aligned}
\ee
As one knows, vortices in models that support the $U(1)$ symmetry usually arise with the presence of the minimal coupling with the gauge field in the derivative $D_\mu=\p_\mu+ieA_\mu$; see Refs.~\cite{NO,jackiw1,jackiw2,coreanos,godvortex}. Here, we deal with generalized models with nonminimal coupling, in which the dual electromagnetic field appears in the derivative, in the new form $\Dc_\mu=\p_\mu+ieA_\mu-iqG(\vphia)F_\mu$, with the function $G(\vphia)$ in principle arbitrary. $P(\vphia)$ denotes a generalized magnetic permeability, $M(\vphia)$ drives the dynamical term of the scalar field and $V(\vphia)$ represents the potential. Despite the general form of the above Lagrange density, the model support a $U(1)$ local symmetry.

We work with natural units $(\hbar=c=1)$ and the dimension of the quantities involved are: $[x^\mu]=\xi^1$, $[\vphi]=[A_\mu]=[e]=\xi^{-\frac12}$, $[\kappa]=\xi^{-1}$, $[q]=\xi^{\frac12}$, $[V(\vphia)]=\xi^{-3}$, where $\xi$ is the dimension of energy. The three functions $G(\vphia)$, $P(\vphia)$ and $M(\vphia)$ are dimensionless.

The equations of motion of the fields $\vphi$ and $A_\mu$ associated to the Lagrange density \eqref{lagrange} are
\bes
\bal\label{eomphi}
&\Dc_\mu\left(M(\vphia)\Dc^\mu\vphi\right) +\frac{\vphi}{2\vphia}\bigg(\frac12P_{\vphia}F_\mu F^\mu -M_{\vphia}\ov{\Dc_\mu\vphi}\Dc^\mu\vphi\nn 
& -\frac{q}{e}G_{\vphia}F_\mu J^\mu +V_{\vphia}\bigg) = 0,\\ \label{eoma}
&\epsilon^{\lambda\mu\nu}\p_\mu\left(P(\vphia)F_\lambda -\frac{q}{e}G(\vphia)J_\lambda\right) -J^\nu +\kappa F^\nu =0,
\eal
\ees
where the current is defined as
\be\label{current}
J_\mu=ieM(\vphia)(\vphic\,\Dc_\mu\vphi-\vphi\,\ov{\Dc_\mu\vphi}),
\ee
and we use the notation $G_{\vphia} = \p G/\p{\vphia}$, $V_{\vphia} = \p V/\p{\vphia}$ and so on. For convenience, we write the fields as
\be\label{Atilde}
\vphi = \vphia e^{i\Lambda} \quad\text{and}\quad A_\mu = \wt{A}_\mu -\frac{1}{e}\p_\mu\Lambda,
\ee
with $\Lambda=\Lambda(x^\mu)$. By doing so, the current in Eq.~\eqref{current} takes the form
\be\label{corrente}
J_\mu = -2e\vphia^2M(\vphia)\left(e\wt{A}_\mu -qG(\vphia)F_\mu\right).
\ee
The energy momentum tensor has the form
\be\label{tmunu}
\begin{aligned}
T_{\mu\nu} &= \left(P -2q^2\vphia^2G^2M\right)\left(F_\mu F_\nu -\frac12\eta_{\mu\nu}F_\lambda F^\lambda\right)\\
	&+M\!\left(2\Re(\ov{D_\mu\vphi}D_\nu\vphi) -\eta_{\mu\nu}\ov{D_\lambda\vphi}D^\lambda\vphi\right) +\eta_{\mu\nu}V,
\end{aligned}
\ee
where $\Re(z)$ denotes the real part of $z$. In particular, the energy density, defined as $\rho\equiv T_{00}$, has the form
\be\label{t00}
\begin{aligned}
\rho &= \left(P -2q^2\vphia^2G^2M\right)\left(F_0^2 -\frac12F_\lambda F^\lambda\right)\\
	&+M\!\left(2|D_0\vphi|^2 -\ov{D_\lambda\vphi}D^\lambda\vphi\right) +V,
\end{aligned}
\ee

To investigate the presence of vortex configurations in the model described by the Lagrange density \eqref{lagrange}, we take static fields and
\be\label{ansatz}
\vphi = g(r)e^{in\theta}, \quad A_0 = h(r) \quad\text{and}\quad \textbf{A} = \frac{\hat{\theta}}{er}(n-a(r)),
\ee
where $(r,\theta)$ are the polar coordinates and $n=\pm1,\pm2,\pm3,...$ is the vorticity. Here, $a(r)$ is dimensionless and $[g(r)]=[h(r)]=\xi^{-\frac12}$. To obtain vortex configurations with finite, single-valued fields at the origin, we impose the boundary conditions
\be
a(0)=n,\quad g(0)=0\quad\text{and}\quad h(0)=h_0,
\ee
where $h_0$ is, in principle, a real finite parameter whose value depends on the specific model. For the functions involved in the transformation \eqref{Atilde}, we must have
\be
\vphia = g(r) \quad\text{e}\quad \wt{\textbf{A}} = -\frac{\hat{\theta}}{er}a(r).
\ee

One may be also interested in the electric field, $\textbf{E}=(E_x,E_y)$, and magnetic field, $B$. For fields in the form \eqref{ansatz}, one can show that $E^i=F^{i0}$ and $B=-F^{12}$ are given by
\be\label{fields}
\textbf{E} = -h^\prime\;\hat{r} \quad\text{and}\quad  B=-\frac{a^\prime}{er}.
\ee
By integrating the above magnetic field, one gets the magnetic flux, which depend on the boundary conditions associated to the specific model defined by $G(\vphia)$, $M(\vphia)$, $P(\vphia)$ and $V(\vphia)$.

The equation of motion \eqref{eomphi} with the fields given by \eqref{ansatz} takes the form
\be\label{eomg}
\begin{aligned}
	&\frac{1}{r}\left(rMg^\prime\right)^\prime +\left(\frac14P_{\vphia} -q^2g^2GMG_{\vphia}\right)\left({h^\prime}^2 -\frac{{a^\prime}^2}{e^2r^2}\right)\\ 
	&+g\left(M +\frac12gM_{\vphia}\right)\left(\left(eh -\frac{qGa^\prime}{er}\right)^2 -\left(\frac{a}{r} -qGh^\prime\right)^2\right)\\
	&+qg^2MG_{\vphia}\left(\frac{ah^\prime}{r} -\frac{ha^\prime}{r}\right) -\frac12\left(M_{\vphia}{g^\prime}^2 +V_{\vphia}\right) = 0.
\end{aligned}
\ee
Similarly, from Eq.~\eqref{eoma}, one gets two equations of motion. They are the version of Gauss' and Amp\`ere's laws for the present model. They are respectively given by
\bes\label{eomaansatz}
\bal\label{gauss}
&\frac{1}{r}\left(\left(P -2q^2g^2G^2M\right)rh^\prime +2qg^2GMa\right)^\prime\\
&-\left(1 -\frac{2qe}{\kappa}g^2GM\right)\frac{\kappa a^\prime}{er} -2e^2g^2Mh = 0,\nn
&\left(\left(P -2q^2g^2G^2M\right)\frac{a^\prime}{er} +2qeg^2GMh\right)^\prime\\
&-\left(1 -\frac{2qe}{\kappa}g^2GM\right)\kappa h^\prime -\frac{2eg^2Ma}{r} = 0.\nonumber
\eal
\ees

The charge density, $J_0$, comes from the definition of the current $J_\mu$ in Eq.~\eqref{current}; it is written as
\be\label{j0}
\begin{split}
J_0 &= -2eg^2M\left(eh -\frac{qGa^\prime}{er}\right) \\
&=\frac{1}{r}\!\left(\!\left(\!1 -\frac{2qe}{\kappa}g^2GM\right)\!\frac{\kappa a}{e} -\left( P -2q^2g^2G^2M\right)\!rh^\prime\!\right)^\prime,
\end{split}
\ee
where we have used Eq.~\eqref{gauss} to get the expression in the latter line. The integration of the above charge density gives the charge of the vortex configuration. It depends on the boundary values of the solutions, which are controlled by $G(\vphia)$, $M(\vphia)$, $P(\vphia)$ and $V(\vphia)$ that define the model.

The energy density with the fields in the form \eqref{ansatz} comes from  Eq.~\eqref{t00}; it is given by
\be\label{dens}
\begin{aligned}
\rho &= \frac12\left(P(g)-2q^2g^2G^2(g)M(g)\right)\left({h^\prime}^2 + \frac{{a^\prime}^2}{e^2r^2}\right)\\
& + M(g)\left(\frac{a^2g^2}{r^2} +e^2g^2h^2+ {g^\prime}^2\right) + V(g).
\end{aligned}
\ee

The equations of motion \eqref{eomg} and \eqref{eomaansatz} that govern the fields are differential equations of  second order with couplings between the functions. So, to simplify the problem, it is of interest to find first order equations compatible with the aforementioned equations. 

\section{First Order Formalism}\label{secthree}

In this Section we focus on the first order formalism, that is, on the presence of first order differential equations that solve the equations of motion of the model. In Refs.~\cite{torres,ghoshplb,ghosh}, the authors found first order equations for models with $M(\vphia)=1$ and specific conditions for $P(\vphia)$ and $G(\vphia)$. Here, in Sec. ref{firstcase} we extend the method to our generalized model described by the Lagrange density in Eq.~\eqref{lagrange}. And later, in Sec. \ref{secondcase} we introduce a novel possibility, which arises under distinct conditions and leads to new first order equations that induce the presence of new vortex configurations.

\subsection{FIRST CASE}\label{firstcase}

The first possibility to find differential equations of the first order type, compatible with the equations of motion \eqref{eomg} and \eqref{eomaansatz}, is to consider a generalization of the trick firstly implemented in \cite{torres} and then generalized in Refs.~\cite{ghoshplb,ghosh}, considering the inclusion of a function that drives the generalized magnetic permeability, with the Lagrange density in the form \eqref{lagrange} under the conditions $M(\vphia)=1$, $P(\vphia)=G(\vphia)$ and $q=e/\kappa$. In this section, we make an extension of his suggestion and take a general $M(\vphia)$ and
\be\label{genghosh}
P(\vphia) = \frac{q\kappa}{e}G(\vphia),
\ee
so $q$ is not constrained to $e$ and $\kappa$. In this case, the Lagrange density in Eq.~\eqref{lagrange} becomes
\be\label{lagrangetrick1}
\begin{aligned}
\LL =& -\frac{q\kappa}{4e}G(\vphia)F_{\mu\nu}F^{\mu\nu} + \frac{\kappa}{4}\epsilon^{\lambda\mu\nu}A_\lambda F_{\mu\nu}\\ 
&+M(\vphia)\ov{\Dc_\mu\vphi}\Dc^\mu\vphi - V(\vphia),
\end{aligned}
\ee
and the equation of motion \eqref{eoma} take the form
\be
\epsilon^{\mu\nu\lambda}\p_\mu\left(\frac{q}{e}G(\vphia)\left(\kappa F_\lambda -J_\lambda\right)\right) + \kappa F^\nu - J^\nu=0.
\ee
This equation is compatible with $J^\mu = \kappa F^\mu$. We remark here that, in this case, the charge density is related to the magnetic field by $J^0 = \kappa F^0=-\kappa B$. Thus, the charge can be written in terms of the magnetic flux, as
\be\label{q1}
Q=-\kappa\Phi.
\ee
By using Eq.~\eqref{Atilde}, we have
\be\label{fmuamu}
\kappa F_\mu = -\frac{2e^2\wt{A}_\mu\vphia^2M(\vphia)}{f(\vphia)},
\ee
with
\be
f(\vphia) = 1-\frac{2qe}{\kappa}\vphia^2G(\vphia)M(\vphia).
\ee
Notice Eq.~\eqref{fmuamu} must be considered with Eq.~\eqref{eomphi} under the condition \eqref{genghosh}, which is of second order. To simplify the problem, we develop the Bogomol'nyi procedure for this case. The expression in Eq.~\eqref{fmuamu} allows us to show that the energy density in Eq.~\eqref{t00} becomes
\be
\begin{aligned}
\rho &= M(\vphia)\Bigg[\left(\p_0\vphia\right)^2 +\left(\p_1\vphia\right)^2 +\left(\p_2\vphia\right)^2\\
	&+\frac{e^2\wt{A}_1^2\vphia^2}{f(\vphia)} +\frac{e^2\wt{A}_2^2\vphia^2}{f(\vphia)}\Bigg] +\frac{\kappa^2F_{0}^2f(\vphia)}{4e^2\vphia^2M(\vphia)} +V(\vphia).
\end{aligned}
\ee
We then introduce the notation $\p_{\pm}=\p_1\pm i\p_2$ e $\wt{A}_{\pm}=\wt{A}_1 \pm i\wt{A}_2$ and write the above expression as
\be
\begin{aligned}
\rho &= M(\vphia)\Bigg(\left(\p_0\vphia\right)^2 +\left|\p_{\pm}\vphia +\frac{ie\tilde{A}_{\pm}\vphia}{\sqrt{f(\vphia)}}\right|^2\Bigg)\\
	&+\frac{\kappa^2f(\vphia)}{4e^2\vphia^2M(\vphia)}\left(F_{0} \pm\frac{2e}{\kappa}\vphia\sqrt{\frac{V(\vphia)M(\vphia)}{f(\vphia)}}\right)^2\\
	& \pm\frac{2e\vphia M(\vphia)\epsilon^{ij}\tilde{A}_j\p_i\vphia}{\sqrt{f(\vphia)}} \mp\frac{\kappa}{e}\sqrt{\frac{V(\vphia)f(\vphia)}{\vphia^2M(\vphia)}}\epsilon^{ij}\p_i\wt{A}_j.
\end{aligned}
\ee
To write the latter three contributions as a single derivative in the above expression, we impose the constraint
\be\label{vinc}
\frac{d}{d\vphia}\left(\sqrt{\frac{V(\vphia)f(\vphia)}{\vphia^2M(\vphia)}}\right) = -\frac{2e^2\vphia M(\vphia)}{\kappa\sqrt{f(\vphia)}}.
\ee
It is satisfied by the potential
\be\label{vvinc}
V(\vphia) = \frac{4e^4\vphia^2M(\vphia)}{\kappa^2f(\vphia)} \left(\int d\vphia\,\frac{\vphia M(\vphia)}{\sqrt{f(\vphia)}}\right)^2.
\ee
Notice we have an indefinite integral in the above expression, so an integration constant will appear in the process. The energy is then given by integrating the energy density in the plane, which we denote by $\Sigma$.
\be
\begin{aligned}
E &=\!\! \int_\Sigma\! d^2x\Bigg(M(\vphia)\Bigg((\p_0\vphia)^2 + \left|\p_{\pm}\vphia +\frac{ie\wt{A}_{\pm}\vphia}{\sqrt{f(\vphia)}}\right|^2\Bigg)\\ 
	&+\frac{\kappa^2f(\vphia)}{4e^2\vphia^2M(\vphia)}\left(F_{0} \pm\frac{2e}{\kappa}\vphia\sqrt{\frac{V(\vphia)M(\vphia)}{f(\vphia)}}\right)^2\\
	& \mp \frac{\kappa}{e}\epsilon^{ij}\p_i\left(\wt{A}_j\sqrt{\frac{V(\vphia)f(\vphia)}{\vphia^2M(\vphia)}}\right)\Bigg).
\end{aligned}
\ee
Since the squared terms in the integral are non negative, the energy is bounded
\be\label{ebgeral1}	E \geq E_B =\frac{\kappa}{e}\left|\int_\Sigma d^2x\,\epsilon^{ij}\p_i\left(\wt{A}_j\sqrt{\frac{V(\vphia)f(\vphia)}{\vphia^2M(\vphia)}}\right)\right|,
\ee
where the potential must obey Eq.~\eqref{vvinc}. If the solutions satisfy the first order equations
\bes\label{fo1geral}
\bal\label{phit}
&\p_0\vphia=0,\\
&\p_{\pm}\vphia +\frac{ie\wt{A}_{\pm}\vphia}{\sqrt{f(\vphia)}} = 0,\\
&F_{0} \pm\frac{2e}{\kappa}\vphia\sqrt{\frac{V(\vphia)M(\vphia)}{f(\vphia)}} = 0,
\eal
\ees
then the energy is minimized to $E=E_B$. The first order equation \eqref{phit}, in particular, is satisfied by static configurations. In this case, $A_0=\wt{A}_0$ and Eq.~\eqref{fmuamu} leads to
\be\label{a01geral}
	A_0 = \pm\frac{1}{e\vphia}\sqrt{\frac{V(\vphia)f(\vphia)}{M(\vphia)}},
\ee
with the potential obeying Eq.~\eqref{vvinc}.
We remark that this first order formalism was developed without requiring $\vphi$ and $A_\mu$ to obey specific expressions. We now show how it works for fields with the specific form in Eq.~\eqref{ansatz}, as it gives a simpler view of the problem. In this case, the equation \eqref{fmuamu} that arises from $J^\mu = \kappa F^\mu$ leads to
\bes\label{fotrick1}
\bal \label{foatrick1}
&\frac{\kappa a^\prime}{er} +\frac{2e^2g^2M(g)h}{f(g)} = 0,\\ \label{htrick1}
&\kappa h^\prime +\frac{2eg^2M(g)a}{rf(g)} = 0,
\eal
\ees
where
\be\label{ftrick1}
f(g) = 1-\frac{2qe}{\kappa}g^2G(g)M(g)
\ee
is a non negative function. The equation of motion for the scalar field \eqref{eomg} becomes
\be\label{eomg1ansatz}
\begin{aligned}
&\frac{1}{r}\left(rMg^\prime\right)^\prime +\frac{g}{f^2}\left(M +\frac12gM_{\vphia} +\frac{qe}{\kappa}g^3M^2G_{\vphia}\right)\\
&\times\left(e^2h^2 -\frac{a^2}{r^2}\right) - \frac12\left(M_{\vphia}{g^\prime}^2 +V_{\vphia}\right) = 0.
\end{aligned}
\ee

By using Eqs.~\eqref{fotrick1} to eliminate $h$ and $h^\prime$, one can show that the energy density in Eq.~\eqref{dens} simplifies to
\be\label{rho}
\rho = M(g)\left({g^\prime}^2 +\frac{a^2g^2}{r^2f(g)}\right) +\frac{\kappa^2f(g){a^\prime}^2}{4e^4r^2g^2M(g)} + V(g).
\ee
So, one must solve Eqs.~\eqref{fotrick1} and \eqref{eomg1ansatz} to calculate the solutions and then substitute them in the above expression to find the corresponding energy density. Nevertheless, we only have two first order equations, Eqs.~\eqref{fotrick1}. Thus, since we have three functions to calculate: $a(r)$, $g(r)$ and $h(r)$, we need an additional first order equation to fulfill our purpose. To solve this issue, we develop the Bogomol'nyi procedure for our model in Eq.~\eqref{lagrangetrick1} under the condition $J^\mu = \kappa F^\mu$. The above energy density can be written in the form
\be
\begin{aligned}
\rho &=M(g)\left(g^\prime \mp\frac{ag}{r\sqrt{f(g)}}\right)^2\\
     &+ \frac{\kappa^2f(g)}{4e^2g^2M(g)}\left(\frac{a^\prime}{er}\pm\frac{2eg}{\kappa}\sqrt{\frac{M(g)V(g)}{f(g)}}\right)^2\\
     &\mp\frac1r\left(\frac{\kappa}{e^2g}\sqrt{\frac{V(g)f(g)}{M(g)}}a^\prime -\frac{2gM(g)}{\sqrt{f(g)}}g^\prime a \right).
\end{aligned}
\ee
To make the latter term become a total derivative, we impose the constraint
\be\label{consttrick1}
\frac{d}{dg}\left(\sqrt{\frac{Vf}{g^2M}}\right) = -\frac{2e^2gM}{\kappa\sqrt{f}}.
\ee
By solving it, one shows that
\be\label{pottrick1}
V(g) =\frac{4e^4g^2M}{\kappa^2f}\left(\int dg\,\frac{gM}{\sqrt{f}}\right)^2,
\ee
where $f$ is as in Eq.~\eqref{ftrick1}. These results are compatible with Eqs.~\eqref{vinc} and \eqref{vvinc}. If the potential has the above form, one can write the energy density as
\be\label{rhosq1}
\begin{aligned}
\rho &=M(g)\left(g^\prime \mp\frac{ag}{r\sqrt{f(g)}}\right)^2 + \frac{\kappa^2f(g)}{4e^2g^2M(g)}\\
     &\times\left(\frac{a^\prime}{er}\pm\frac{2eg}{\kappa}\sqrt{\frac{M(g)V(g)}{f(g)}}\right)^2\pm\frac1r W^\prime,
\end{aligned}
\ee
where $W=W(a,g)$ is an auxiliary function given by
\be\label{wtrick1}
\begin{split}
    W(a,g) &= -\frac{\kappa a }{e^2g}\sqrt{\frac{V(g)f(g)}{M(g)}}\\
           &= 2a\,\int dg\,\frac{gM(g)}{\sqrt{f(g)}}.
\end{split}
\ee
We have used Eq.~\eqref{pottrick1} to get the above Eq. \eqref{wtrick1}. Since the integral of the energy density in Eq.~\eqref{rhosq1} gives the energy, one can see the energy is bounded
\be\label{ebtrick1}
	E\geq E_B = 2\pi\left| W(a(\infty),g(\infty)) - W(a(0),g(0))\right|.
\ee
 Notice that, differently from what occurs in the general procedure Eq.~\eqref{ebgeral1}, we can show there is a surface term that gives the energy, given by $W(a,g)$.

The configurations with minimum energy appear when we take the squared terms equal to zero. In this case, we get the first order equations
\bes\label{fo1}
\bal\label{fo1g}
g^\prime &= \pm \frac{ag}{r}\frac{1}{\sqrt{f}},\\ \label{fo1a}
-\frac{a^\prime}{er} &= \pm \frac{2eg}{\kappa}\sqrt{\frac{MV}{f}}\\
                     &= \mp \frac{4e^3g^2M}{\kappa^2f}\left(\int dg\,\frac{gM}{\sqrt{f}}\right).\nonumber
\eal
\ees
We emphasize here the presence of the function $f$ in Eq.~\eqref{fo1g}. In models with minimal coupling, we only get this equation with $f=1$, i.e., $g^\prime=ag/r$, which only leads to an integer power law behavior near the origin for $g(r)$ (see Ref.~\cite{godvortex}). Since we now have a general $f$ in the form \eqref{ftrick1}, we may consider functions $G(g)$ and $M(g)$ to obtain distinct behaviors around the origin. Furthermore, the above first order equation \eqref{fo1a} allows us to conclude that $a(r)$ is a monotonically decreasing/increasing function for the upper/lower sign. For topological solutions, we have that $g(r)$ is monotonically increasing, connecting $g=0$ and $g=v$, and both $a(r)$ and $a^\prime(r)$ does not change its sign, oppositely as we will see in the models of the next section. In the case of nontopological solutions, the sign $a^\prime(r)$ is constant, but $a(r)$ changes along its path, so $g(r)$ increases up to a maximum value and then smoothly decreases towards $g=0$.

If the solutions obey these first order equations, the energy is minimized to $E=E_B$, with $E_B$ given by Eq.~\eqref{ebtrick1}. By solving the above equations, one can find $h$ through Eq.~\eqref{foatrick1} combined with the latter equation given above. We then have
\be\label{h1}
\begin{split}
	h &= \pm\frac{1}{eg}\sqrt{\frac{Vf}{M}}\\
	  &= \mp \frac{2e}{\kappa}\left(\int dg\,\frac{gM}{\sqrt{f}}\right).
\end{split}
\ee
The equations \eqref{fo1} and \eqref{h1} are compatible with Eqs.~\eqref{fo1geral} and \eqref{a01geral}. For models with $G(g)$, $M(g)$ and $V(g)$ obeying the constraint \eqref{consttrick1}, i.e., for potentials in the form \eqref{pottrick1}, the equations \eqref{fo1} and the above one completely solve the problem. To find $a(r)$ and $g(r)$, one must solve the first order equations \eqref{fo1}. The remaining solution, $h(r)$, is found by substituting $g(r)$ in Eq.~\eqref{h1}; it has the same sign of the vorticity. We note that the upper signs represent configurations with positive vorticity and the lower ones do it for negative vorticity. They are related by the changes $a(r)\to-a(r)$ and $h(r)\to-h(r)$; $g(r)$ remains the same in both scenarios. For simplicity, we only work with positive vorticity here.

We now illustrate our procedure with a generalization of the models considered by in \cite{torres} and  \cite{ghoshplb,ghosh}, given by
\be\label{g1}
G(g) = \frac{\kappa}{2qe}\frac{1- M^2(g)\left(1 -\alpha g^2\right)^{1-\gamma}}{g^2 M(g)},
\ee
where $\alpha$ and $\gamma$ are parameters such that $\alpha$ has the dimension of energy and $\gamma$ is dimensionless. From Eq.~\eqref{ftrick1}, we obtain $f(g) = M^2(g)\left(1 -\alpha g^2\right)^{1-\gamma}$. Notice that, in principle, $M(g)$ is arbitrary, restricted only by the non-negative character of the energy density. For the above function, by taking $M=1$, $\alpha=1/v^2$ and $\gamma=0$, one recovers the model in Ref.~\cite{torres}, in which $G(g) = \kappa/(2qev^2)$. On the other hand, by taking $M=1$ and $\alpha=1/v^2$, one gets the model in Refs.~\cite{ghoshplb,ghosh}, where $G(g)=\kappa(1-(1-g^2/v^2)^{1-\gamma})/(2qeg^2)$. For a general $M(g)$, the above equation can be substituted in Eq.~\eqref{pottrick1} to obtain the potential
\be\label{v1c}
V(g) = \frac{4e^4g^2\left(1 -\alpha g^2\right)^{\gamma -1}}{\kappa^2\alpha^2\left(1+\gamma\right)^2M(g)}\left(C -\left(1 -\alpha g^2\right)^{\frac{\gamma +1}{2}}\right)^2.
\ee
Here, $C$ is an integration constant. Assuming $M(g)$ does not modify the minimum $g=v$ of the potential, we have to be careful with the sign of $\alpha$. We deal with $\alpha>0$, by taking $\alpha=1/v^2$. In this case, the above potential does not need a $C\neq0$ to support symmetry breaking. So, for simplicity, we take $C=0$. By doing this, the above expression simplifies, becoming
\be\label{v1c0}
V(g) =\frac{4e^4v^4g^2}{\kappa^2\left(1+\gamma\right)^2M(g)}\left(1 -\frac{g^2}{v^2}\right)^{2\gamma }.
\ee
To find the solutions, one must solve the first order equations \eqref{fo1}, which, in this model, can be written as
\bes\label{fo11}
\bal
g^\prime &= \frac{ag}{rM(g)}\left(1 -\frac{g^2}{v^2}\right)^{\frac{\gamma-1}{2}},\\
-\frac{a^\prime}{er} &= \frac{4e^3v^2g^2}{\kappa^2(1+\gamma)M(g)}\left(1 -\frac{g^2}{v^2}\right)^{\frac{3\gamma -1}{2}}.
\eal
\ees
Notice one must suggest a function $M(g)$ to obtain the solutions. The function $h(r)$ in Eq.~\eqref{h1} is given in terms of the known $g(r)$ as
\be\label{h11}
h(r)=\frac{2ev^2}{\kappa\,(1+\gamma)}\,\left(1-\frac{g^2(r)}{v^2}\right)^\frac{\gamma+1}{2}.
\ee
At the origin, since $g(0)=0$, we have $h(0)= 2ev^2/\left(\kappa(1+\gamma)\right)$. Note that $h$ does not depend explicitly on $M(g)$. However, $M(g)$ modifies the profile of $g(r)$, which must be substituted in the above equation. Moreover, the electric field depends on $M(g)$, since it is given by Eq.~\eqref{fields}.

The auxiliary function in Eq.~\eqref{wtrick1} takes the form
\be\label{w11}
W(a,g) = -\frac{2v^2a}{(1+\gamma)}\left(1 -\frac{g^2}{v^2}\right)^{\frac{\gamma +1}{2}}.
\ee
So, the energy does not depend on $M(g)$, since it depends only on the boundary values of $a(r)$ and $g(r)$, with $E=E_B$, where $E_B$ is as in Eq.~\eqref{ebtrick1}. On the other hand, the energy density is modified by $M(g)$, because this function changes the solutions $a(r)$ and $g(r)$ such that the energy density is changed. This can be straightforwardly seen by writing the energy density in Eq.~\eqref{rho} in terms of the solutions $a(r)$ and $g(r)$, in the form
\be\label{rhofields11}
\begin{aligned}
\rho &= \frac{2g^2}{M(g)}\left(1-\frac{g^2}{v^2}\right)^{\gamma-1}\\
     &\times\left(\frac{a^2}{r^2} + \frac{4e^4v^4}{\kappa^2\left(1+\gamma\right)^2}\left(1-\frac{g^2}{v^2}\right)^{\gamma+1} \right),
\end{aligned}
\ee
where we have used the first order equations \eqref{fo11}.

Notice that one must solve Eqs.~\eqref{fo11} and substitute the solutions in the above equation to calculate the energy density. However, one must be careful to take the appropriate boundary conditions as they are related to the topological or nontopological nature of the vortex configurations. So, we review the simplest case, which appears for $M=1$ and was studied in Refs.~\cite{torres,ghoshplb,ghosh}. In particular, in Ref.~\cite{torres}, as $M=1$ and $\gamma=0$, the energy density becomes
\be
\rho = 2g^2\left(\frac{a^2}{r^2}\left(1-\frac{g^2}{v^2}\right)^{-1} + \frac{4e^4v^4}{\kappa^2} \right).
\ee
Hence, to ensure the finiteness of the energy, one must impose $a(\infty) \to a_\infty$ and $g(\infty)\to0$. This means that the vortex solutions found by Torres engender nontopological character. The model investigated in \cite{torres} was generalized in Refs.~\cite{ghoshplb,ghosh}, with the inclusion of a generalized magnetic permeability driven by the scalar field. For $M=1$ and $\gamma=1$, one recovers the pure Chern-Simons model, with the energy density given by 
\be
\rho = 2g^2\left(\frac{a^2}{r^2} + \frac{e^4v^4}{\kappa^2}\left(1-\frac{g^2}{v^2}\right)^{2}\right),
\ee
which was studied in Refs.~\cite{jackiw1,jackiw2,coreanos} and may lead to topological ($a(\infty)\to0$ and $g(\infty)\to v$), or non topological ($a(\infty)\to a_\infty$ and $g(\infty)\to0$) solutions. For $M=1$ and $\gamma>1$, one can see from Eq.~\eqref{rhofields11} that topological solutions are supported by the model, i.e., $a(\infty)\to a_\infty$ and $g(\infty)\to v$. Surprisingly, the very same model support nontopological solutions, $a(\infty)\to a_\infty$ and $g(\infty)\to0$, since the global factor $g^2$ goes to zero and protects the energy against divergences. By using Eqs.~\eqref{w11} and \eqref{ebtrick1} one can show that, in this case, the topological solutions has energy $E=4\pi v^2n/(1+\gamma)$, whilst the nontopological solutions engender energy $E=4\pi v^2(n-a_\infty)/(1+\gamma)$.

We now show that our model with $M\neq1$ support vortex configurations. Considering the function $G(g)$ in Eq.~\eqref{g1}, one can take, for instance, 
\be\label{m0}
M(g) = \left(1-\frac{g^2}{v^2}\right)^{\frac{\sigma-1}{2}},
\ee
where $\sigma$ is a dimensionless parameter. This makes the first order equations \eqref{fo11} become
\bes
\bal
g^\prime &= \frac{ag}{r}\left(1 -\frac{g^2}{v^2}\right)^{\frac{\gamma-\sigma}{2}},\\
-\frac{a^\prime}{er} &= \frac{4e^3v^2g^2}{\kappa^2(1+\gamma)}\left(1 -\frac{g^2}{v^2}\right)^{\frac{3\gamma -\sigma}{2}}.
\eal
\ees
Notice that, for $\sigma\neq \gamma$, the above equations are similar to the ones found in Refs.~\cite{ghoshplb,ghosh}. A particular case is $\sigma=\gamma$, in which $G(g)=0$, such that the Maxwell and the dual field $F^\mu$ term in the derivative $\mathcal{D}$ that governs the non-minimal coupling vanish, so the gauge and scalar fields are minimally coupled and we get a model that falls in the class of generalized pure Chern-Simons models investigated in Ref.~\cite{yang}.

As we commented before, the function $M(g)$ in Eq.~\eqref{m0} support solutions with similar behavior to the ones found in Refs.~\cite{ghoshplb,ghosh}. Next, we introduce a function that leads to novel vortex configurations, with
\be
M(g) = \left(\lambda +(1-\lambda)\frac{g^2}{v^2}\right)\left(1 -\frac{g^2}{v^2}\right)^{\frac{\gamma -1}{2}},
\ee
where $\lambda$ is a dimensionless parameter such that $\lambda\in(0,1]$. Notice that $\lambda=1$ recovers Eq.~\eqref{m0} with $\sigma=\gamma$. For a general $\lambda$, we get from Eq.~\eqref{g1} that
\be
G(g) = \frac{\kappa}{2qe}\frac{1-\left(\lambda +(1-\lambda)g^2/v^2\right)^2}{g^2\left(\lambda +(1-\lambda)g^2/v^2\right)}\left(1 -\frac{g^2}{v^2}\right)^{\frac{1-\gamma}{2}}.
\ee
One must take into account that this function also drives the generalized magnetic permeability, $P(g)$, since they are related as in Eq.~\eqref{genghosh}. We remark that, differently from what occurs in the models in Refs.~\cite{torres,ghoshplb,ghosh}, the above function is non-negative in the interval where the topological solutions exists, $g\in[0,v]$. Notice $\lambda=\gamma=1$ recovers the standard pure Chern-Simons model \cite{jackiw1,jackiw2,coreanos}. The potential in Eq.~\eqref{v1c0} takes the form
\be
V(g) = \frac{4e^4v^4g^2}{\kappa^2\left(1 +\gamma\right)^2}\left(\lambda +(1-\lambda)\frac{g^2}{v^2}\right)^{-1}\left(1 -\frac{g^2}{v^2}\right)^{\frac{3\gamma +1}{2}}.
\ee
Its minima are located at $g=0$ and $g=v$. The first order equations \eqref{fo11} are
\bes\label{first1}
\bal
\label{first1a}g^\prime &= \frac{ag}{r}\left(\lambda +(1-\lambda)\frac{g^2}{v^2}\right)^{-1},\\
-\frac{a^\prime}{er} &= \frac{4e^3v^2g^2}{\kappa^2(1+\gamma)}\left(\lambda +(1-\lambda)\frac{g^2}{v^2}\right)^{-1}\left(1 -\frac{g^2}{v^2}\right)^\gamma.
\eal
\ees
We now investigate the behavior of the solutions near the origin. For $r\approx0$, one can take $a(r)=n-a_{o}(r)$ and $g(r)=g_{o}(r)$ for small $a_{o}$ and $g_{o}$ to show that the above equations lead to 
\be
a_{o}(r)\propto r^{\frac{2(n+\lambda)}{\lambda}} \quad\text{and}\quad g_{o}(r)\propto r^{\frac{n}{\lambda}}.
\ee
Notice that the presence of $\lambda$ in the power of these functions occurs due to the factor $\left(\lambda +(1-\lambda)g^2/v^2\right)^{-1}$ in Eq.~\eqref{first1a}. So, since $\lambda$ is a real positive parameter, we get a real number in the power of $r$. To find $h(r)$, one must solve the above first order equations and substitute the known $g(r)$ in Eq.~\eqref{h11}. The energy density in Eq.~\eqref{rhofields11} becomes
\be\label{dens1}
\begin{aligned}
\rho &= 2g^2\left(\lambda +(1-\lambda)\frac{g^2}{v^2}\right)^{-1}\left(1-\frac{g^2}{v^2}\right)^{\frac{\gamma-1}{2}}\\
     &\times\left(\frac{a^2}{r^2} + \frac{4e^4v^4}{\kappa^2\left(1+\gamma\right)^2}\left(1-\frac{g^2}{v^2}\right)^{\gamma+1} \right).
\end{aligned}
\ee
We see from the above equation that the model support finite energy nontopological and topological solutions for $\gamma\geq1$. Here, we only deal with topological solutions, such that $a(\infty)=0$ and $g(\infty)=v$. From Eqs.~\eqref{w11} and \eqref{ebtrick1}, it is straightforward to show that the energy of the topological solutions is given by $E=4\pi|n|v^2/(1+\gamma)$. Moreover, by integrating the magnetic field in Eq.~\eqref{fields}, one can show the flux is $\Phi = {2\pi n}/{e}$, so both the magnetic flux and the charge in Eq.~\eqref{q1} are quantized. 

\begin{figure}[t!]
\centering
\includegraphics[width=4.2cm,trim={0.6cm 0.2cm 0 0},clip]{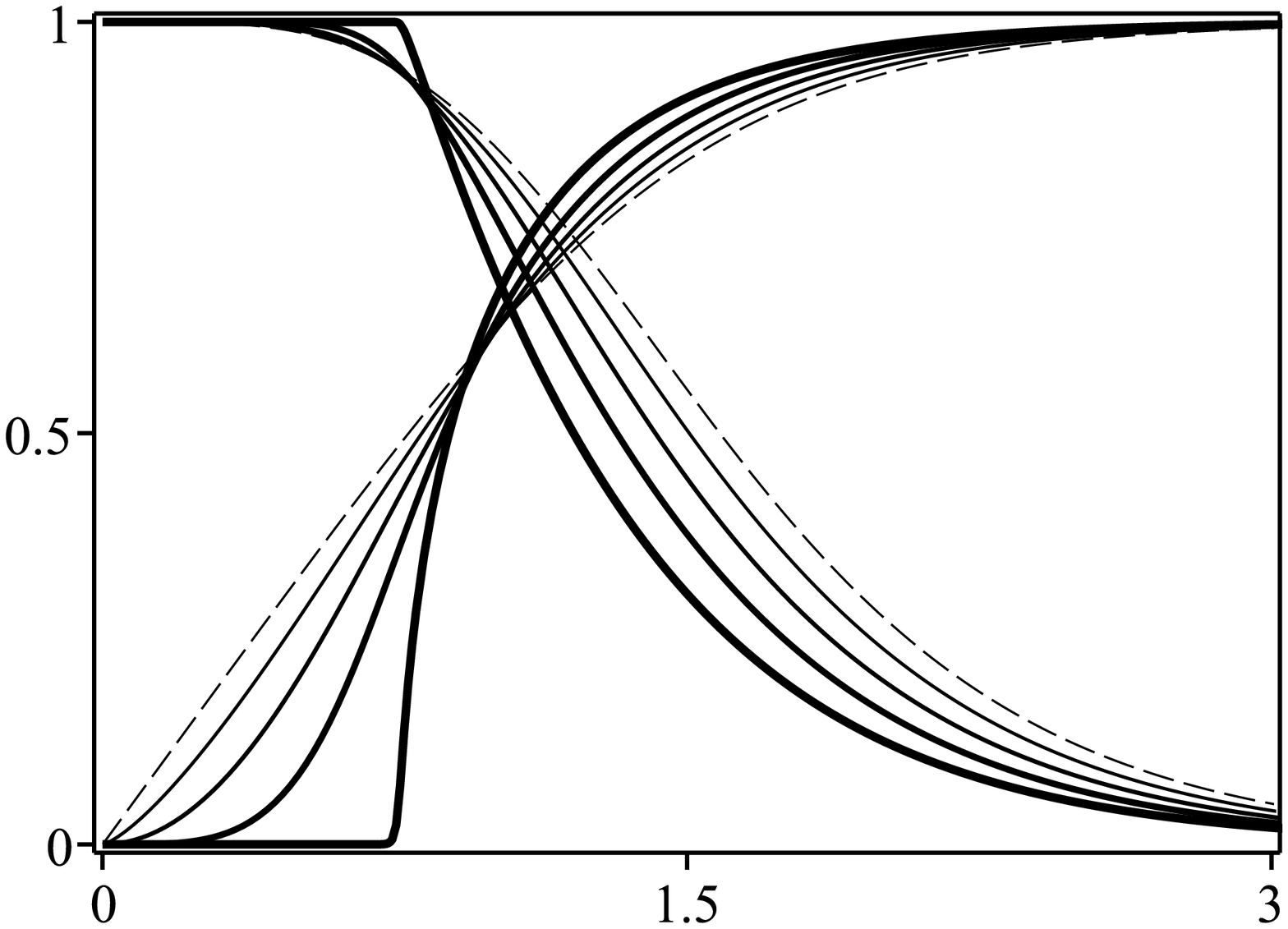}
\includegraphics[width=4.2cm,trim={0.6cm 0.2cm 0 0},clip]{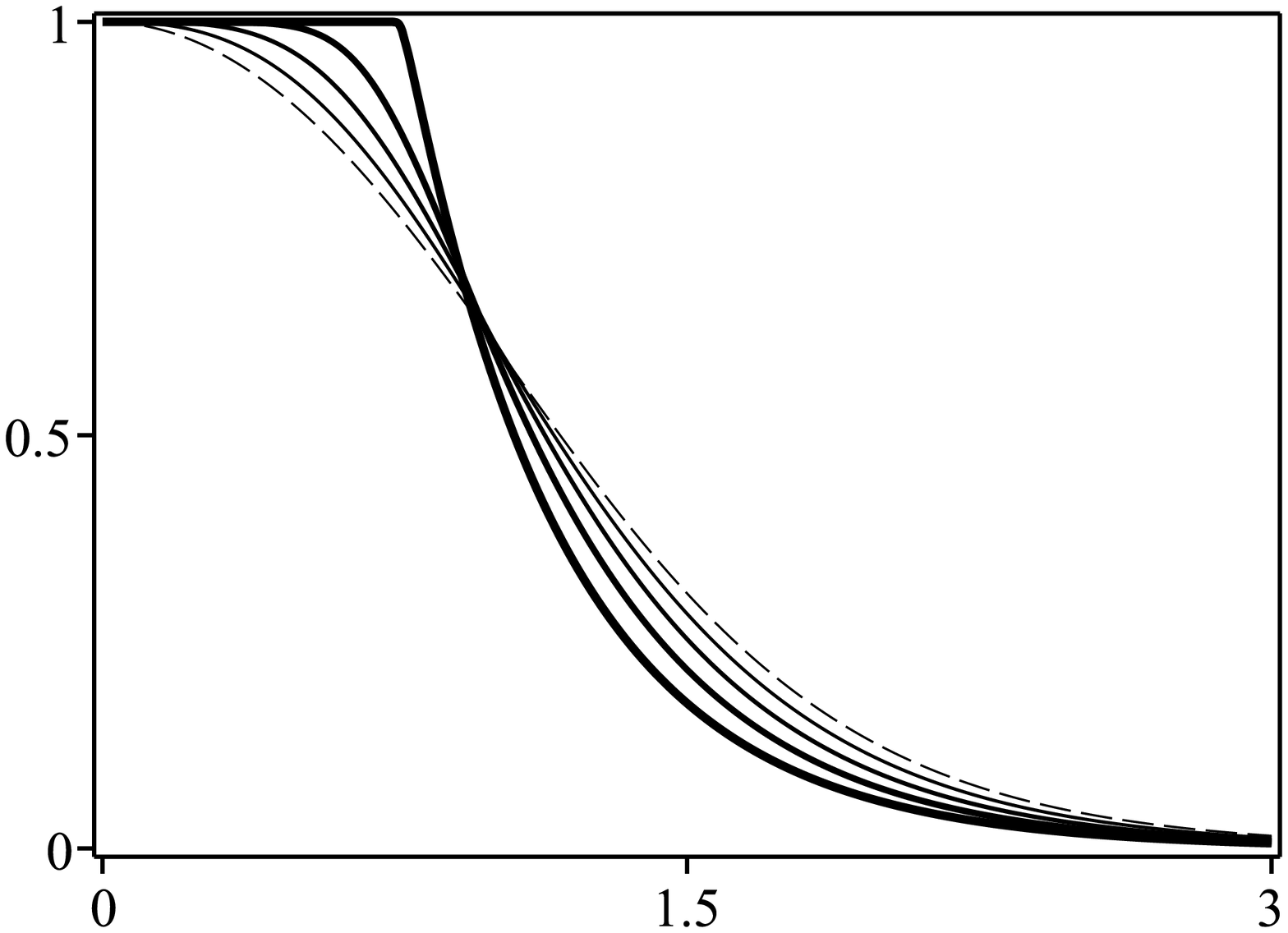}
\caption{The solutions $a(r)$ and $g(r)$ of Eq.~\eqref{first1} (left) and the function $h(r)$ in Eq.~\eqref{h11} (right) for $e=\kappa=q=v=n=\gamma=1$ and $\lambda=0.01,0.25,0.5,0.75$ and $1$. The thickness of the lines decreases with $\lambda$, and the dashed line represents $\lambda=1$, which is the pure Chern-Simons model.}
\label{fig1}
\end{figure}

\begin{figure}[t!]
\centering
\includegraphics[width=4.2cm,trim={0.6cm 0.2cm 0 0},clip]{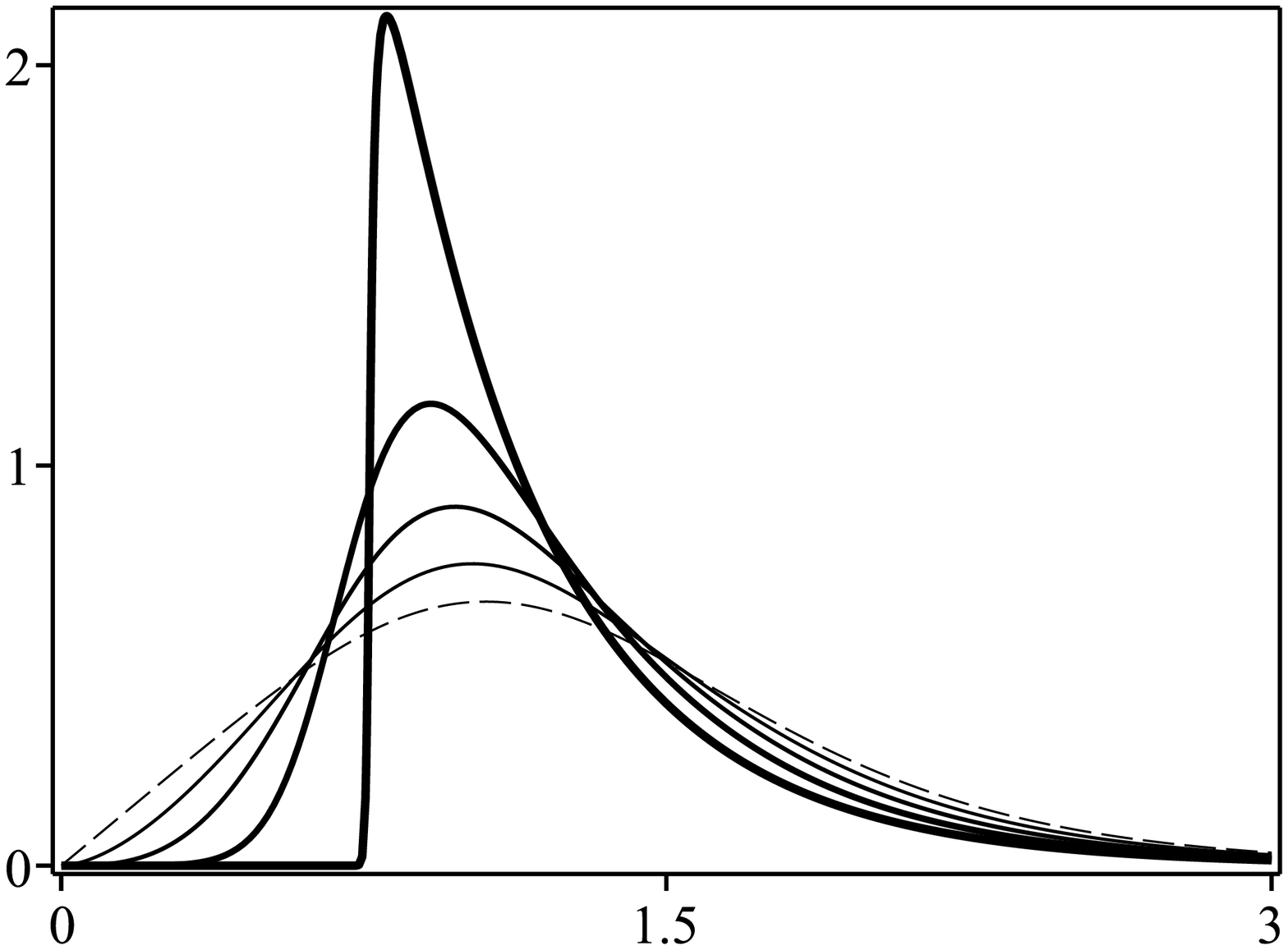}
\includegraphics[width=4.2cm,trim={0.6cm 0.2cm 0 0},clip]{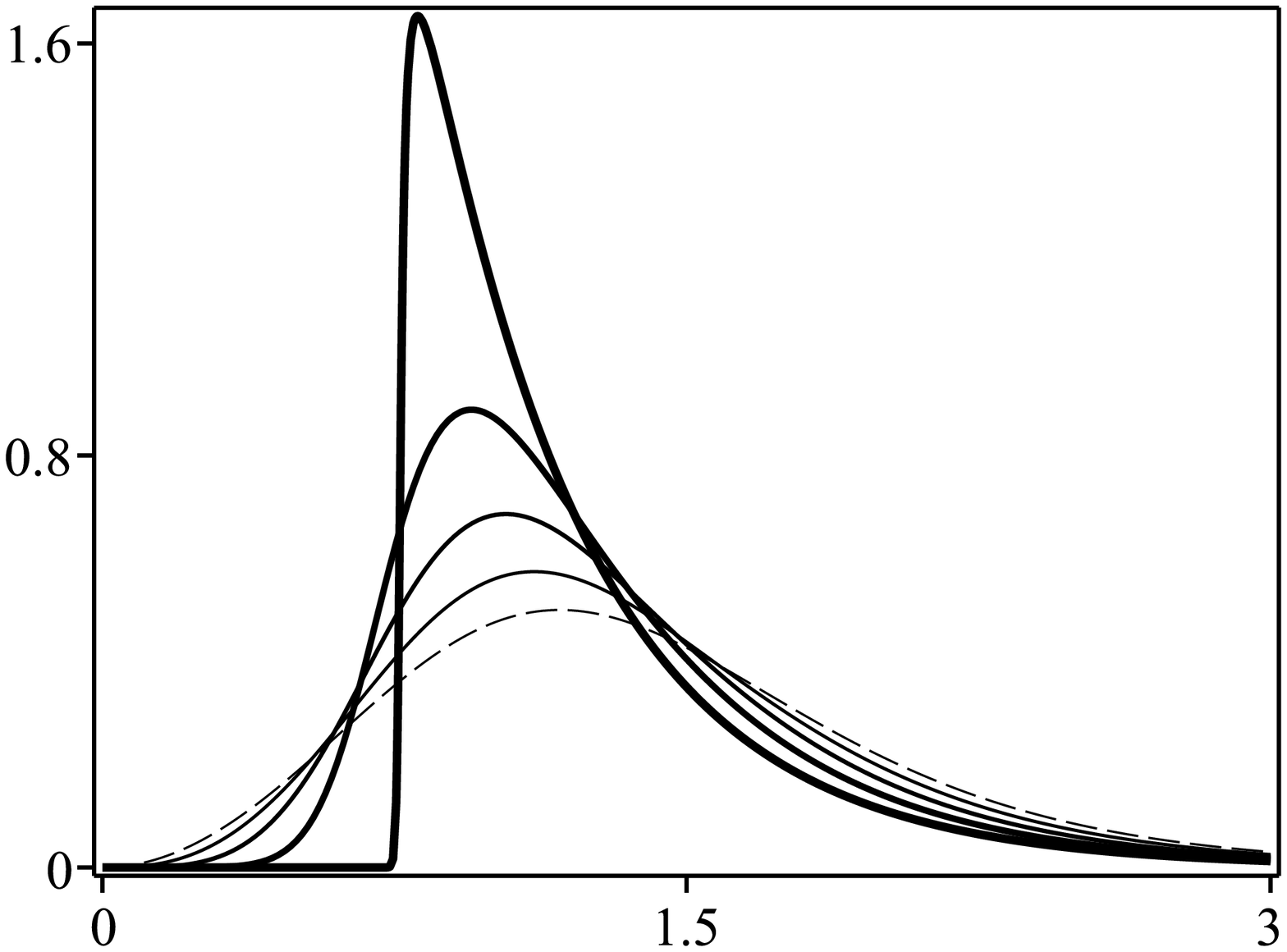}
\caption{The intensity of the electric (left) and the magnetic (right) fields in Eq.~\eqref{fields} for the solutions of Eq.~\eqref{first1} with $e=\kappa=q=v=n=\gamma=1$ and $\lambda=0.01, 0.25,0.5,0.75$ and $1$. The thickness of the lines decreases with $\lambda$, and the dashed line represents $\lambda=1$, which is the pure Chern-Simons model.}
\label{fig2}
\end{figure}

\begin{figure}[t!]
\centering
\includegraphics[width=6cm,trim={0.6cm 0.2cm 0 0},clip]{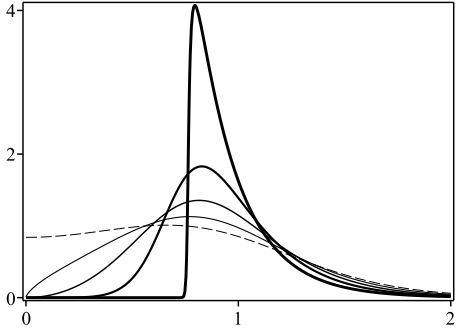}
\caption{The energy density in Eq.~\eqref{dens1} for the solutions of Eq.~\eqref{first1} with $e=\kappa=q=v=n=\gamma=1$ and $\lambda=0.01, 0.25,0.5,0.75$ and $1$. The thickness of the lines decreases with $\lambda$, and the dashed line represents $\lambda=1$, which is the pure Chern-Simons model.}
\label{fig3}
\end{figure}

We then use numerical procedures to solve the involved first order equations and find the profiles of $a(r)$, $g(r)$ and $h(r)$. They can be seen in Fig.~\ref{fig1}, where we plot these functions for $e=\kappa=q=v=n=\gamma=1$ and some values of $\lambda$. One can see that, as $\lambda$ decreases, the plateau that appears in each one of the solutions becomes wider. Notice that, even though the solutions behave distinctively near the origin, the tail of the solutions are very similar for the several values of $\lambda$. By using these solutions, we calculate the associated electric and magnetic fields, and the energy density. These quantities are displayed in Figs.~\ref{fig2} and \ref{fig3}. Notice that they have a hole around the origin, whose deepness and width become larger as $\lambda$ decreases.

Notice the results in Figs.~\ref{fig1}, \ref{fig2} and \ref{fig3} are for $\alpha=1/v^2>0$ and $C=0$ in Eq.~\eqref{v1c}. One may also obtain well defined vortex configurations for $C\neq0$; this will be explored in the next section, in which we present a new manner to develop a first order formalism for models described by the Lagrange density in Eq.~\eqref{lagrange}.

\subsection{SECOND CASE}\label{secondcase}
In the previous section, we have dealt with the first order formalism that can be developed under the condition $J^\mu = \kappa F^\mu$ for the Lagrange density in Eq.~\eqref{lagrange}, a generalization of the models investigated in Refs.~\cite{torres,ghoshplb,ghosh}. We, however, have found a distinct pathway that leads us to first order equations compatible with the equations of motion \eqref{eomg} and \eqref{eomaansatz} for the model in Eq.~\eqref{lagrange}. Instead of constraining $P(\vphia)$ and $G(\vphia)$ as in the previous case, the trick here is to take
\be\label{trick2}
P(\vphia) = 2q^2{\vphia}^2G^2(\vphia)M(\vphia),
\ee
which makes the Lagrange density in Eq.~\eqref{lagrange} become
\be\label{lagrangetrick2}
\begin{aligned}
\LL &= -\frac{q^2}{2}\vphia^2G^2(\vphia)M(\vphia)F_{\mu\nu}F^{\mu\nu} \\ 
&+ \frac{\kappa}{4}\epsilon^{\lambda\mu\nu}A_\lambda F_{\mu\nu} +M(\vphia)\ov{\Dc_\mu\vphi}\Dc^\mu\vphi - V(\vphia).
\end{aligned}
\ee
One can expand the above expression to show that the Maxwell term vanishes and the Lagrangian density can be written as
\be
\begin{aligned}
\LL &= \frac{\kappa}{4}\epsilon^{\lambda\mu\nu}\!\!\left(\!A_\lambda \!+\!\frac{2iq}{\kappa}G(\vphia)M(\vphia)\!\left(\vphic D_\lambda\vphi -\vphi\ov{D_\lambda\vphi}\right)\!\right)\!F_{\mu\nu}\\
	&+M(\vphia)\ov{D_\mu\vphi}D^\mu\vphi - V(\vphia),
\end{aligned}
\ee
which is a generalization of the Chern-Simons model investigated in Ref.~\cite{burzlaff}.

The equation of motion for the gauge field \eqref{eoma} with Eq.~\eqref{Atilde} becomes
\be\label{fmuamu2}
\kappa\epsilon^{\lambda\mu\nu}\p_\mu\left(\wt{A}_\lambda\sqrt{f(\vphia)}\right) = \frac{2e^2\wt{A}^\nu\vphia^2M(\vphia)}{\sqrt{f(\vphia)}},
\ee
where
\be\label{fcase2}
f(\vphia) = 1 -\frac{4qe}{\kappa}\vphia^2G(\vphia)M(\vphia).
\ee
To calculate the solutions, one must solve Eq.~\eqref{fmuamu2} and \eqref{eomphi} under the condition \eqref{trick2}. Even though Eq.~\eqref{fmuamu2} is of first order, one can see that Eq.~\eqref{eomphi} is of second order. In order to get first order equations that describe the system, we develop the Bogomol'nyi procedure, similarly as it was done in the previous case in Sec.~\ref{firstcase}. The condition \eqref{trick2} makes the energy density in Eq.~\eqref{t00} to become
\be
\rho = M(\vphia)\left(2|D_0\vphi|^2 -\ov{D_\lambda\vphi}D^\lambda\vphi\right) +V(\vphia).
\ee
By setting $\nu=0$ in Eq.~\eqref{fmuamu2}, one gets an expression for $\wt{A}_0$, which we use in the above equation to obtain
\be
\begin{aligned}
\rho =& M(\vphia)\Big[\left(\p_0\vphia\right)^2 +\left(\p_1\vphia\right)^2 +\left(\p_2\vphia\right)^2\\
	&+e^2\wt{A}_1^2\vphia^2 +e^2\wt{A}_2^2\vphia^2\Big]+\frac{\kappa^2f(\vphia)}{4e^2\vphia^2M(\vphia)}\\
	&\times\left(\epsilon^{ij}\p_i\left(\wt{A}_j\sqrt{f(\vphia)}\right)\right)^2 +V(\vphia).
\end{aligned}
\ee
It can be rewritten in the form
\be
\begin{aligned}
\rho &= M(\vphia)\bigg(\left(\p_0\vphia\right)^2 +\left|\p_{\pm}\vphia +ie\wt{A}_{\pm}\vphia\right|^2\bigg)\\
	& +\frac{\kappa^2f(\vphia)}{4e^2\vphia^2M(\vphia)}\Bigg(\epsilon^{ij}\p_i\left(\wt{A}_j\sqrt{f(\vphia)}\right)\\
	& \pm\frac{2e}{\kappa}\vphia\sqrt{\frac{V(\vphia)M(\vphia)}{f(\vphia)}}\Bigg)^2\pm2e\vphia M(\vphia)\epsilon^{ij}\wt{A}_j\p_i\vphia\\
	& \mp\frac{\kappa}{e}\sqrt{\frac{V(\vphia)f(\vphia)}{\vphia^2M(\vphia)}}\epsilon^{ij}\p_i\left(\wt{A}_j\sqrt{f(\vphia)}\right).
\end{aligned}
\ee
By using the constraint in Eq.~\eqref{vinc} with $f(\vphia)$ given as in Eq.~\eqref{fcase2}, we have
\be
\begin{aligned}
E &= \int_\Sigma d^2x\Bigg(M(\vphia)\bigg(\left(\p_0\vphia\right)^2 + \left|\p_{\pm}\vphia +ie\wt{A}_{\pm}\vphia\right|^2\bigg)\\ 
	&+\frac{\kappa^2f(\vphia)}{4e^2\vphia^2M(\vphia)}\Bigg(\epsilon^{ij}\p_i\left(\wt{A}_j\sqrt{f(\vphia)}\right) \\
	&\pm\frac{2e}{\kappa}\vphia\sqrt{\frac{V(\vphia)M(\vphia)}{f(\vphia)}}\Bigg)^2 \\
	&\mp \frac{\kappa}{e}\epsilon^{ij}\p_i\left(\wt{A}_jf(\vphia)\sqrt{\frac{V(\vphia)}{\vphia^2M(\vphia)}}\right)\Bigg).
\end{aligned}
\ee
Similarly to the previous case investigated in Sec.~\ref{firstcase}, the energy is bounded, that is,
\be\label{eb2geral}
E \geq E_B = \frac{\kappa}{e}\left|\int_\Sigma d^2x\,\epsilon^{ij}\p_i\left(\wt{A}_jf(\vphia)\sqrt{\frac{V(\vphia)}{\vphia^2M(\vphia)}}\right)\right|.
\ee
If the fields satisfy the first order equations
\bes
\bal\label{foa0case2}
&\p_0\vphia=0,\\
&\p_{\pm}\vphia +ie\wt{A}_{\pm}\vphia = 0,\\
&\p_i\left(\epsilon^{ij}\tilde{A}_j\sqrt{f(\vphia)}\right) \pm\frac{2e}{\kappa}\vphia\sqrt{\frac{V(\vphia)M(\vphia)}{f(\vphia)}} = 0,
\eal
\ees
then the energy is minimized to $E=E_B$. To comply with Eq.~\eqref{foa0case2}, we consider static configurations. In this case, we get from Eq.~\eqref{fmuamu2} that
\be\label{a02geral}
A_0 = \pm\frac{1}{e\vphia}\sqrt{\frac{V(\vphia)}{M(\vphia)}}.
\ee
Notice the results in Eqs.~\eqref{fmuamu2}-\eqref{a02geral} are obtained without suggesting the form of the fields. We then consider the fields in the form \eqref{ansatz}. This makes the charge density \eqref{j0} with the condition in Eq.~\eqref{trick2} to be
\be\label{j02}
J_0=\frac{\kappa}{er}\left(\left(1 -\frac{2qe}{\kappa}g^2GM\right)a\right)^\prime.
\ee
So, it is not proportional to the magnetic field as the case in the previous section, and we cannot ensure the charge is related to the magnetic flux. Since the charge depends on the boundary conditions, we will work it out in the examples.

To get a clearer view of the procedure, we develop the first order formalism with the fields in the form \eqref{ansatz}. In this case, Gauss' and  Amp\`ere's laws are given by Eq.~\eqref{fmuamu2}, that is
\bes\label{fotrick2b}
\bal\label{fotrick2ba}
&\frac{\kappa fa^\prime}{er} -\frac{2qa}{r}\left(g^2GM\right)_gg^\prime +2e^2g^2Mh = 0,\\ 
&\kappa fh^\prime -2qeh\left(g^2GM\right)_gg^\prime +\frac{2eg^2Ma}{r} = 0,
\eal
\ees
where $f$ is henceforth given by the expression
\be\label{ftrick2}
f(g) = 1- \frac{4qe}{\kappa}g^2G(g)M(g).
\ee
We emphasize this definition is different from the one of the previous case, in Eq.~\eqref{ftrick1}. Note the equations of motion \eqref{fotrick2b} related to the gauge field are of first order. Nevertheless, they are not enough to solve the problem since we must calculate three functions: $a(r)$, $g(r)$ and $h(r)$. The third equation is given by \eqref{eomg}, which simplifies to
\be
\begin{aligned}
&\frac{1}{r}\left(rMg^\prime\right)^\prime +\frac{g}{f}\left(M +\frac12gM_{\vphia} +\frac{2qe}{\kappa}g^3M^2G_{\vphia}\right)\\
&\times\left(e^2h^2 -\frac{a^2}{r^2}\right) - \frac12\left(V_{\vphia} +M_{\vphia}{g^\prime}^2\right) = 0.
\end{aligned}
\ee
Notice, however, that the above equation of motion is of second order. Thus, we need to find additional conditions to obtain first order equations that solves the problem. Similarly as in case investigated in the previous section, we develop the Bogomol'nyi procedure here. To do so, we write the energy density in Eq.~\eqref{dens} for the constraint in Eq.~\eqref{trick2}:
\be
\rho = M(g)\left(e^2g^2h^2 + {g^\prime}^2 + \frac{a^2g^2}{r^2}\right) + V(g).
\ee
To eliminate $h$, we use Eq.~\eqref{fotrick2ba} in the above equation to get
\be\label{rho2}
\rho \!=\! M(g)\!\left(\!{g^\prime}^2 \!+\!\frac{a^2g^2}{r^2}\!\right)\! +\frac{\kappa^2f(g)}{4e^4r^2g^2M(g)}{\left(\!a\sqrt{f(g)}\right)^\prime}^2\!\! +\! V(g).\,
\ee
One can show the above equation can be written as
\be
\begin{aligned}
\rho &=M(g)\left(g^\prime \mp\frac{ag}{r}\right)^2\\
     &+ \frac{\kappa^2f(g)}{4e^2g^2M(g)}\left(\frac{1}{er}\left(a\sqrt{f(g)}\right)^\prime\pm\frac{2eg}{\kappa}\sqrt{\frac{M(g)V(g)}{f(g)}}\right)^2\\
     &\mp\frac1r\left(\frac{\kappa}{e^2g}\sqrt{\frac{V(g)f(g)}{M(g)}}\left(a\sqrt{f(g)}\right)^\prime -2gM(g)g^\prime a \right).
\end{aligned}
\ee
To make the latter term become a total derivative, we impose the constraint in Eq.~\eqref{consttrick1}, whose solution is given by the potential in Eq.~\eqref{pottrick1}. We emphasize, however, that the function $f$ is now given by Eq.~\eqref{ftrick2}. So, one may think that the potential here must have the same form of the one in the previous section, but this is not true because the function $f$ is different. In this case, we can write the energy density as
\be\label{rhowlinha}
\begin{aligned}
\rho &=M(g)\left(g^\prime \mp\frac{ag}{r}\right)^2+ \frac{\kappa^2f(g)}{4e^2g^2M(g)}\\
     &\times\left(\frac{1}{er}\left(a\sqrt{f(g)}\right)^\prime\pm\frac{2eg}{\kappa}\sqrt{\frac{M(g)V(g)}{f(g)}}\right)^2 \pm\frac1r W^\prime,
\end{aligned}
\ee
where $W=W(a,g)$ is an auxiliary function given by
\be\label{wtrick2}
\begin{split}
    W(a,g) &= -\frac{\kappa a f(g)}{e^2g}\sqrt{\frac{V(g)}{M(g)}}\\
           &= 2a\sqrt{f(g)}\left(\int dg\,\frac{gM(g)}{\sqrt{f(g)}}\right),
\end{split}
\ee
here we have used the expression in Eq.~\eqref{pottrick1} for the potential. By integrating the energy density in Eq.~\eqref{rhowlinha}, one gets that the energy is bounded exactly as in Eq.~\eqref{ebtrick1} with $W$ given as in the above equation. Notice the function $W(a,g)$ is associated to a surface term that comes from the integration and appears only for fields in the form \eqref{ansatz}, differently from the general procedure in Eq.~\eqref{eb2geral}. The energy is minimized to $E=E_B$ if the following first order equations are satisfied
\bes\label{fo2}
\bal\label{fo2g}
g^\prime &= \pm\frac{ag}{r},\\ \label{fo2a}
-\frac{a^\prime}{er} &= \pm \frac{g}{f}\left(\frac{f_ga^2}{2er^2} +\frac{2e}{\kappa}\sqrt{MV}\right)\\
                     &= \pm \frac{g}{f}\left(\frac{f_ga^2}{2er^2} -\frac{4e^3gM}{\kappa^2\sqrt{f}}\left(\int dg\,\frac{gM}{\sqrt{f}}\right)\right).\nonumber
\eal
\ees
Notice the above first order equation \eqref{fo2g} is different from \eqref{fo1g} since its right side presents only a linear term in $g$. Surprisingly, Eq.~\eqref{fo2g} is the very same which arises in the first order formalism for vortices in models with minimal coupling \cite{godvortex}, so the function $g(r)$ near the origin is always a power-law function, in the form
\be\label{gori}
g(r)\propto r^{|n|}.
\ee
On the other hand, the first order equation \eqref{fo2a} presents two terms, with one of them depending on $a^2$, oppositely from Eq.~\eqref{fo1a}, in which the right side shows only $g$ and $r$. Depending on the sign of $f_g$, the term in $a^2$ may compete with the other one. This is interesting, because, as we will show next, brings novel configurations to light; it does not appear in the usual approach taken in Refs.~\cite{torres,ghoshplb,ghosh} nor in models with minimal coupling in the form investigated in Ref.~\cite{godvortex}, in which $a'/(er)$ is equal to a function of $g$. Furthermore, since the behavior of $g(r)$ near the origin is given by the above equation, one can show that, in this regime, the function $f$ must behave as $f_g/f\propto g^{m/|n|-1}$, with $m>2$, to make the magnetic field $B=-a^\prime/(er)$ be finite. The function $h(r)$ is obtained from Eq.~\eqref{fotrick2ba}
\be\label{h2}
\begin{split}
	h &= \pm \frac{1}{eg}\sqrt{\frac{V}{M}}\\
	  &= \mp \frac{2e}{\kappa\sqrt{f}}\left(\int dg\,\frac{gM}{\sqrt{f}}\right).
\end{split}
\ee
We emphasize this first order formalism is compatible with Eqs.~\eqref{fmuamu2}-\eqref{a02geral}. In the above equations \eqref{fo2} and \eqref{h2}, the upper/lower sign describes configurations with positive/negative vorticity. One can relate these possibilities by making the changes $a(r)\to-a(r)$ and $h(r)\to-h(r)$. For simplicity, we only deal with positive vorticity.

Since we are now dealing with novel first order equations, we investigate a simple model, with
\be\label{mgtrick2}
G(g)=\frac{\kappa}{4qe}\frac{1 -M^2(g)\left(1-\alpha g^2\right)^{1-\gamma}}{g^2M(g)},
\ee
so the function that controls the magnetic permeability is given by Eq.~\eqref{trick2}, which reads
\be
P(g) = \frac{\kappa^2}{8e^2}\frac{\left(1 -M^2(g)\left(1-\alpha g^2\right)^{1-\gamma}\right)^2}{g^2M(g)}.
\ee
Here, $\alpha$ is a parameter with the dimension of energy and $\gamma\geq1$ is a dimensionless parameter. In this model, one obtains from Eq.~\eqref{ftrick2} that $f(g)=M^2(g)\left(1-\alpha g^2\right)^{1-\gamma}$. This makes the potential in Eq.~\eqref{pottrick1} be written as
\be\label{v2c}
V(g) = \frac{4e^4g^2\left(1 -\alpha g^2\right)^{\gamma -1}}{\kappa^2\alpha^2\left(1+\gamma\right)^2M(g)}\left(C -\left(1 -\alpha g^2\right)^{\frac{\gamma +1}{2}}\right)^2,
\ee
where $C$ is an integration constant. Similarly to the case investigated in the previous section, we first consider $C=0$ and $\alpha=1/v^2$. Notice the specific model investigated in Ref.~\cite{burzlaff} is recovered for $M(g)={(1-g^2/v^2)}^2$ and $\gamma=3$. We then consider a new model, with
\be
M(g) = \left(1 +\lambda\,\frac{g^2}{v^2}\right)^{\frac{\sigma}{2}}\!\left(1 -\frac{g^2}{v^2}\right)^{\frac{\gamma-1}{2}},
\ee
where $\lambda$ is a non negative dimensionless parameter. The above potential in Eq.~\eqref{v2c} simplifies to
\be\label{pot21}
V(g) = \frac{4e^4v^4g^2}{\kappa^2\left(1+\gamma\right)^2}\left(1 +\lambda \,\frac{g^2}{v^2}\right)^{-\frac{\sigma}{2}}\!\left(1-\frac{g^2}{v^2}\right)^{\frac{3\gamma+1}{2}}.
\ee
Its minima are located at $g=0$ and $g=v$. We must solve Eqs.~\eqref{fo2} with the upper signs. Note, however, that the first order equation \eqref{fo2g} does not change with $M(g)$ and $G(g)$. On the other hand, the first order equation \eqref{fo2a} takes the form
\be\label{first21a}
\begin{aligned}
-\frac{a^\prime}{er} &= \frac{\sigma\lambda a^2g^2}{ev^2r^2}\left(1 +\lambda \,\frac{g^2}{v^2}\right)^{-1} \\
                     &+ \frac{4e^3v^2g^2}{\kappa^2\left(\gamma+1\right)}\left(1 +\lambda\,\frac{g^2}{v^2}\right)^{-\sigma}\!\left(1-\frac{g^2}{v^2}\right)^\gamma.
\end{aligned}
\ee
By knowing the solutions, one can calculate $h(r)$ from Eq.~\eqref{h2}, which leads us to
\be\label{h21}
h(r)= \frac{2ev^2}{\kappa\left(1+\gamma\right)}\left(1 +\lambda \,\frac{g^2(r)}{v^2}\right)^{-\frac{\sigma}{2}}\!\left(1-\frac{g^2(r)}{v^2}\right)^{\frac{\gamma+1}{2}}.
\ee
At the origin $r=0$, since $g(0)=0$, we have $h(0)=2ev^2/\left(\kappa(1+\gamma)\right)$. To check if this function has critical points outside the origin, we take the derivative of the above expression with respect to $r$. Since $g^\prime>0$, we take $h_g=0$, which leads us to
\be
\tilde{g}=v\sqrt{\frac{1}{\lambda}\frac{\lambda\sigma+\gamma+1}{\sigma-(\gamma+1)}}.
\ee
If $0<\tilde{g}<v$, the solution $h(r)$ presents a critical point. By using this argument, one can show that this function supports a global maximum for $\sigma<0$ and $\lambda>\lambda_c$, with
\be\label{lambdac}
\lambda_c=\frac{\gamma+1}{|\sigma|},
\ee
which leads to an internal structure in the the electric field. The auxiliary function $W(a,g)$ is calculated from Eq.~\eqref{wtrick2}; it has the form
\be\label{w21}
W(a,g) = -\frac{2v^2a}{\left(1+\gamma\right)}\left(1 +\lambda \,\frac{g^2}{v^2}\right)^{\frac{\sigma}{2}}\!\left(1-\frac{g^2}{v^2}\right)^{\frac{\gamma+1}{2}}.
\ee
The energy density is calculated from Eq.~\eqref{rho2}. We make use of the first order equations \eqref{fo2g} and \eqref{first21a} to get the expression
\be\label{dens21}
\begin{aligned}
\rho &= 2g^2\left(1 -\frac{g^2}{v^2}\right)^{\frac{\gamma-1}{2}}\left(1 +\lambda\,\frac{g^2}{v^2}\right)^{-\frac{\sigma}{2}}\\
	&\times\left(\frac{a^2}{r^2}\!\left(1 +\lambda\,\frac{g^2}{v^2}\right)^\sigma + \frac{e^4v^4}{\kappa^2\left(1+\gamma\right)^2}\!\left(1-\frac{g^2}{v^2}\right)^{\gamma+1}\right).
\end{aligned}
\ee
To find the above energy density, one must solve the first order equations \eqref{fo2g} and \eqref{first21a}, and then substitute the solutions in the above expression. In order to calculate the solutions, though, one must be careful with the boundary conditions, which are associated to the topological character of the vortex. From the above equation, we see that the energy is finite for topological solutions, in which $a(\infty)\to 0$ and $g(\infty)\to v$, and also for nontopological solutions, with $a(\infty)\to a_\infty$ and $g(\infty)\to 0$.

For simplicity, we only calculate topological solutions here. By using Eqs.~\eqref{w21} and \eqref{ebtrick1}, one can use the aforementioned boundary conditions to show their energy is $E=4\pi |n|v^2/(1+\gamma)$. Moreover, by integrating the magnetic field in Eq.~\eqref{fields}, one can see that the flux associated to topological solutions is $\Phi=2\pi n/e$. The charge density in Eq.~\eqref{j02} becomes
\be\label{j021}
J_0 = \frac{\kappa}{2er}\left(a+a\left(1+\lambda\,\frac{g^2}{v^2}\right)^\sigma\right)^\prime.
\ee
By integrating it, one can show that $Q=-2\pi n\kappa/e$. Notice both $Q$ and $E$ are quantized and also, even though we have $J_0\neq-\kappa B$, the charge is proportional to the flux: $Q=-\kappa \Phi$.

\begin{figure}[t!]
\centering
\includegraphics[width=4.2cm,trim={0.6cm 0.2cm 0 0},clip]{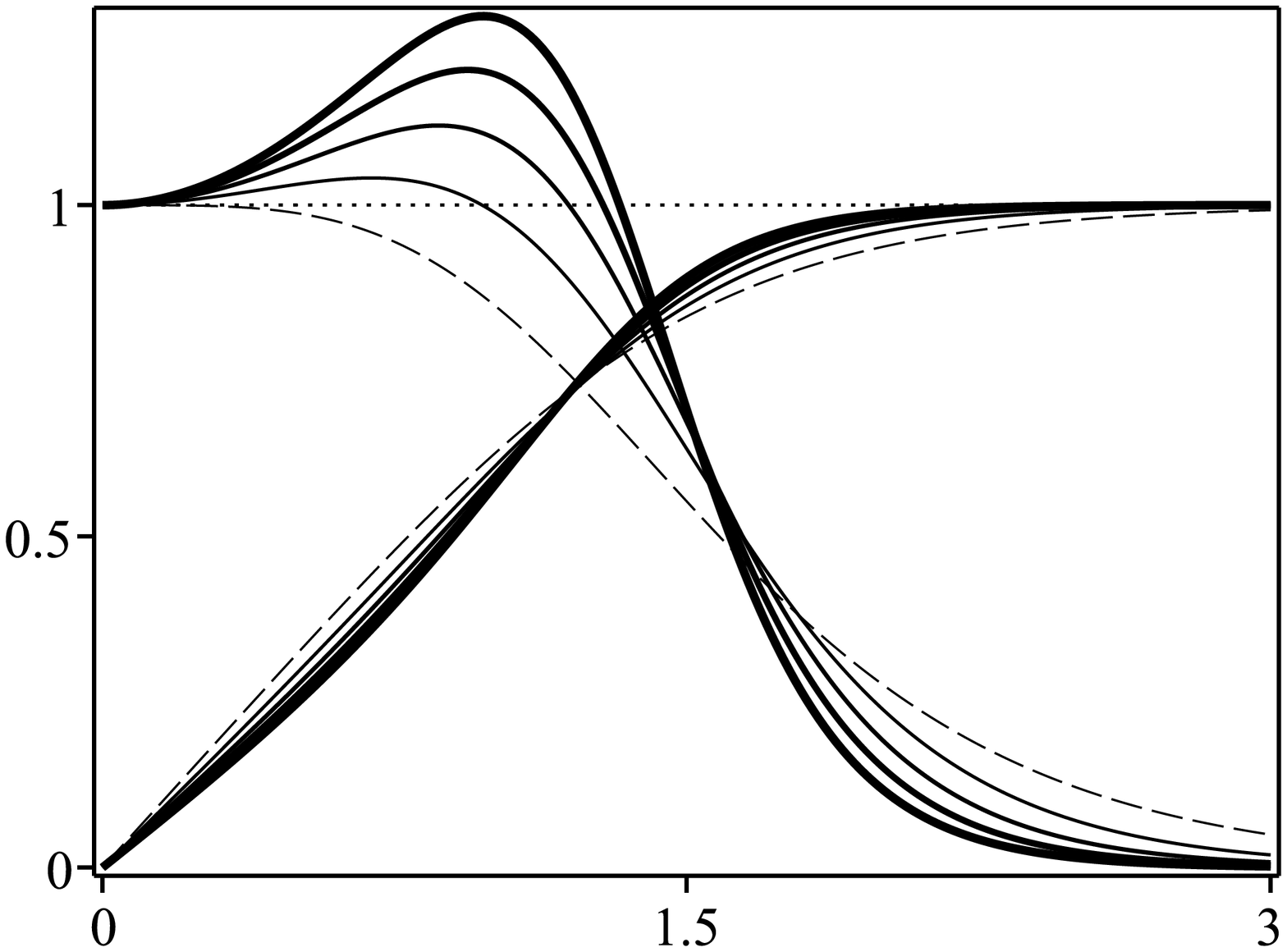}
\includegraphics[width=4.2cm,trim={0.6cm 0.2cm 0 0},clip]{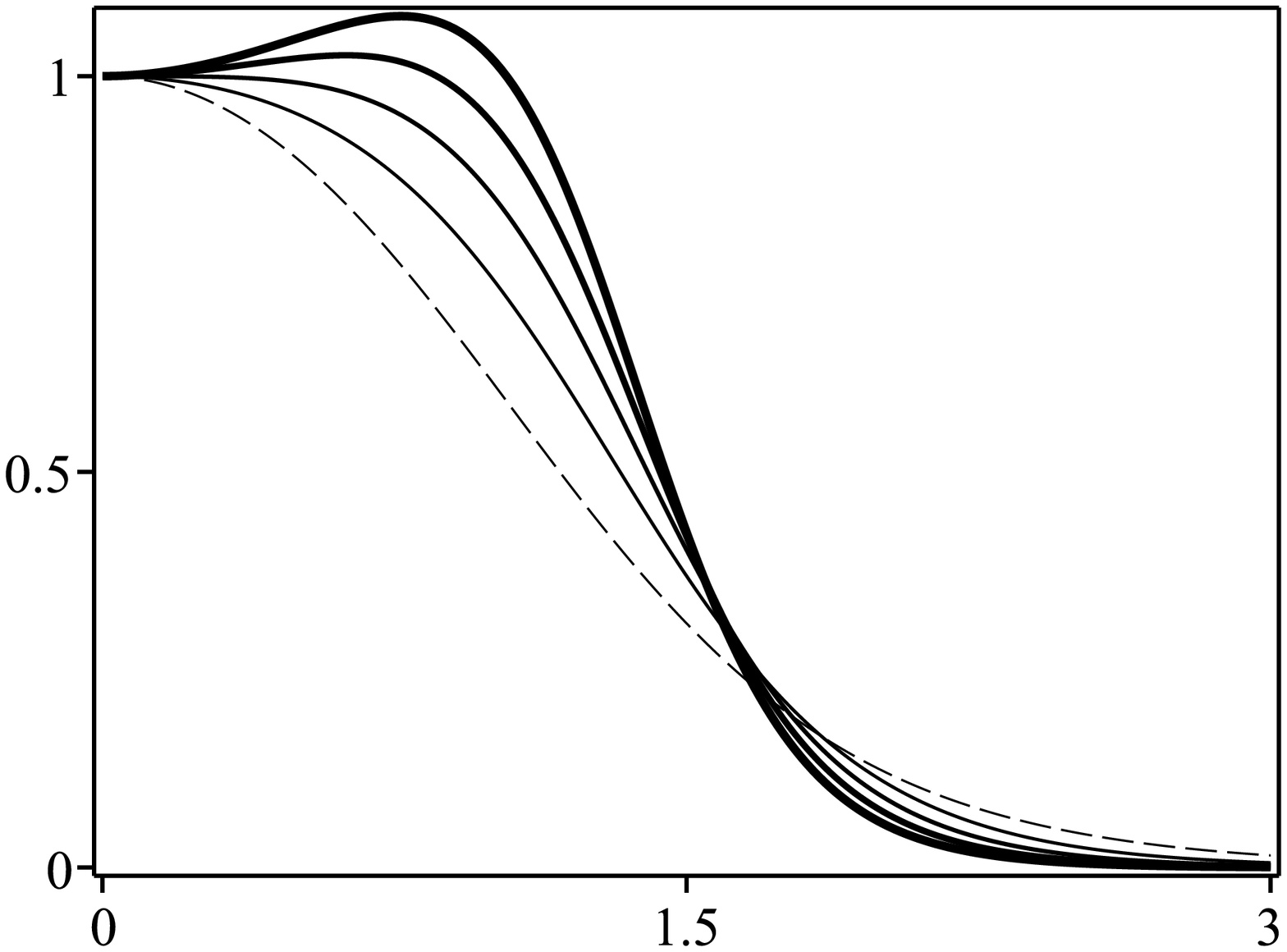}
\includegraphics[width=4.2cm,trim={0.6cm 0.2cm 0 0},clip]{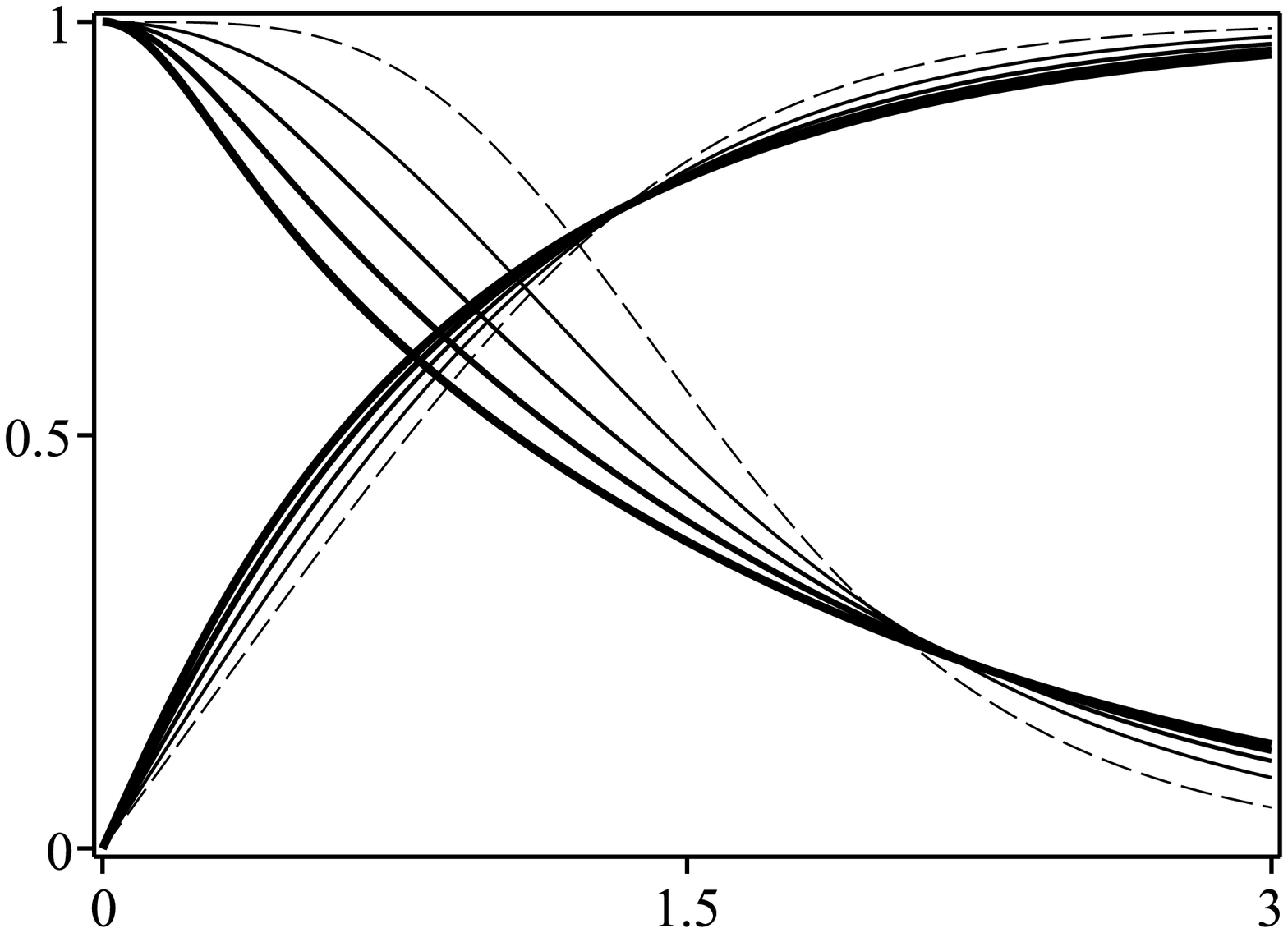}
\includegraphics[width=4.2cm,trim={0.6cm 0.2cm 0 0},clip]{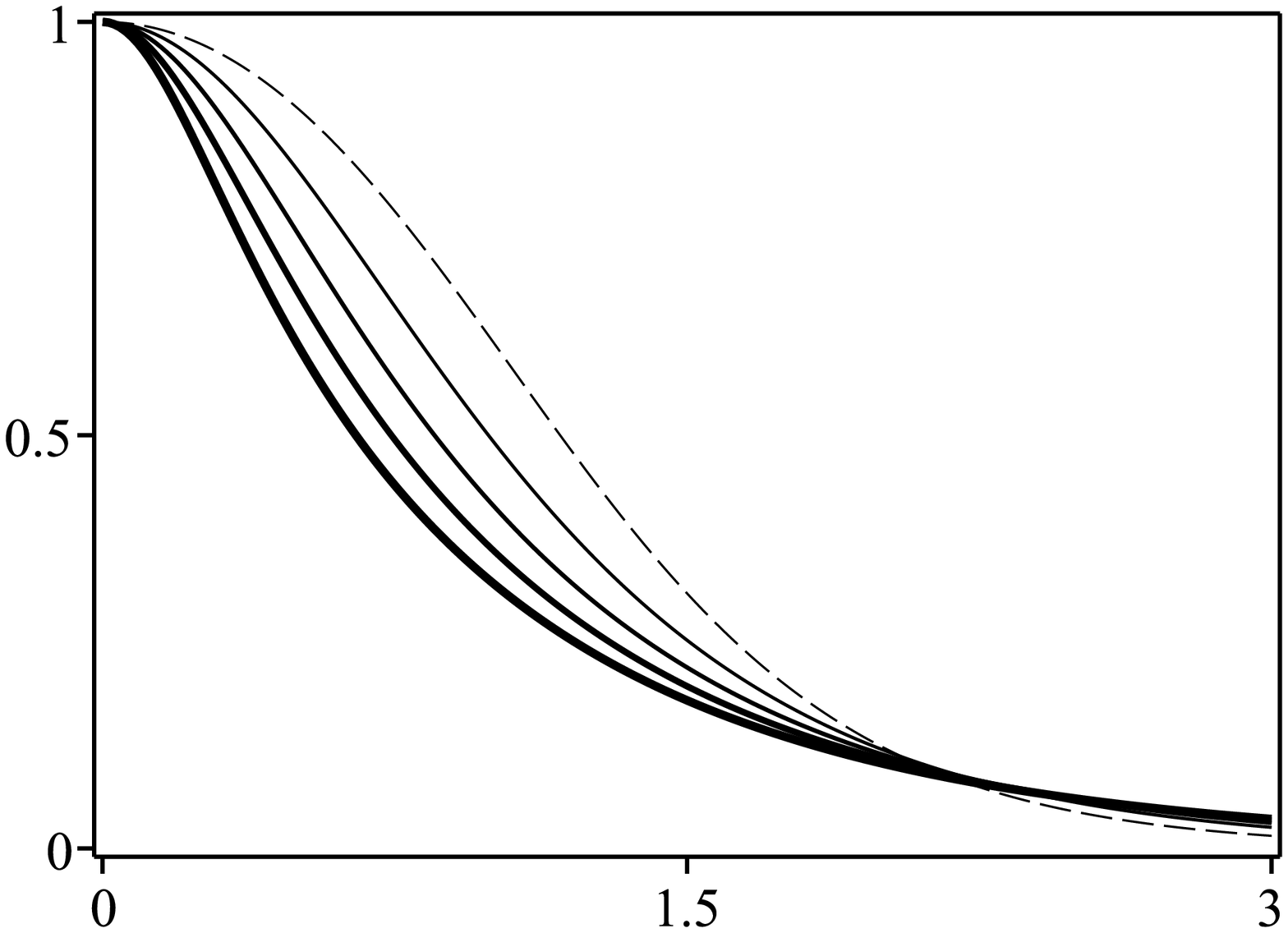}
\caption{The solutions $a(r)$ and $g(r)$ of Eqs.~\eqref{fo2g} and \eqref{first21a} (left) and the function $h(r)$ in Eq.~\eqref{h21} (right) for $e=\kappa=q=v=n=\gamma=1$, $\sigma=-1$ (top) and $1$ (bottom), $\lambda=1,2,3,4$. The dashed lines represent the case $\lambda=0$ and thickness of the lines increases with $\lambda$.}
\label{fig4}
\end{figure}

\begin{figure}[t!]
\centering
\includegraphics[width=4.2cm,trim={0.6cm 0.2cm 0 0},clip]{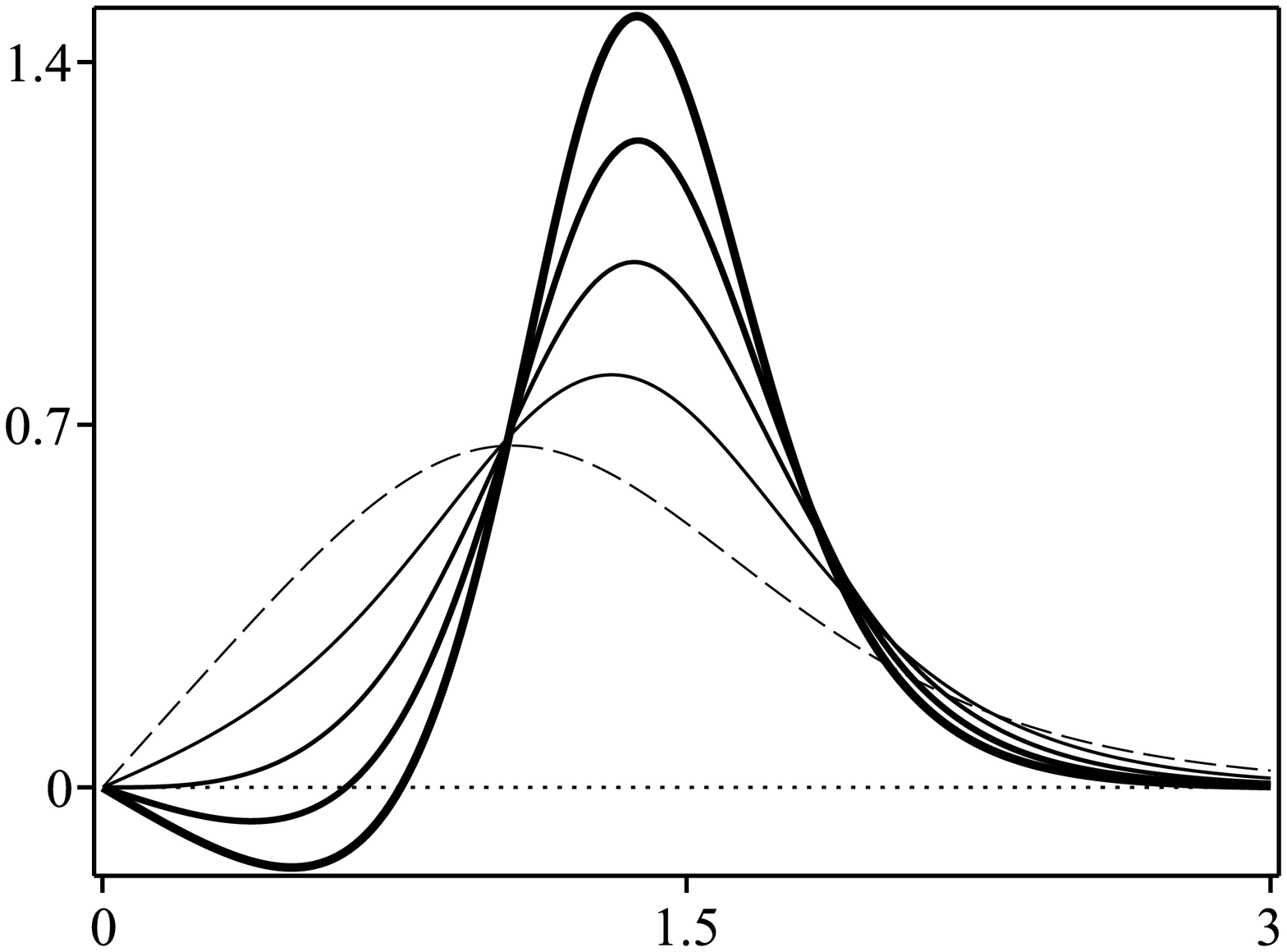}
\includegraphics[width=4.2cm,trim={0.6cm 0.2cm 0 0},clip]{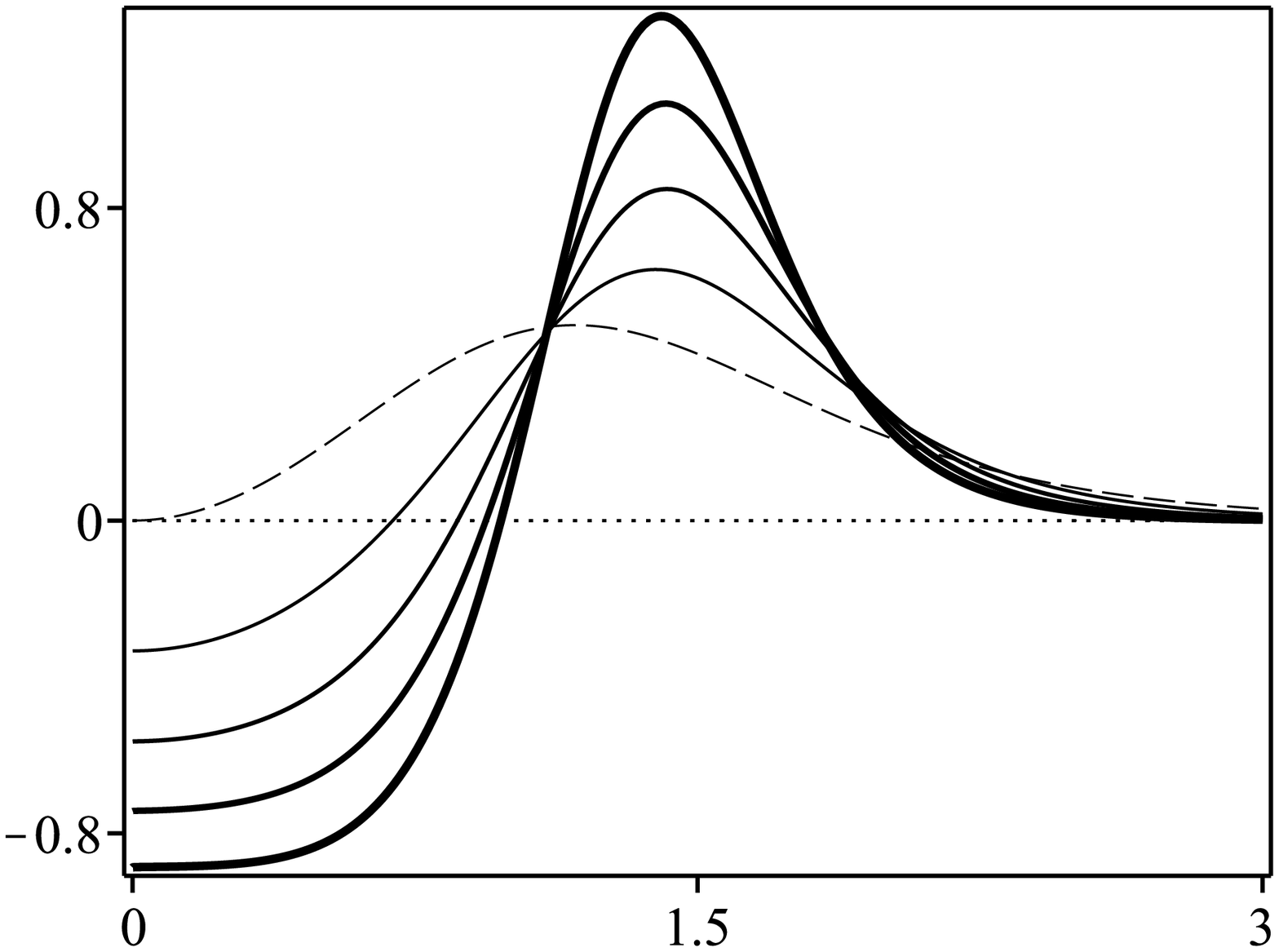}
\includegraphics[width=4.2cm,trim={0.6cm 0.2cm 0 0},clip]{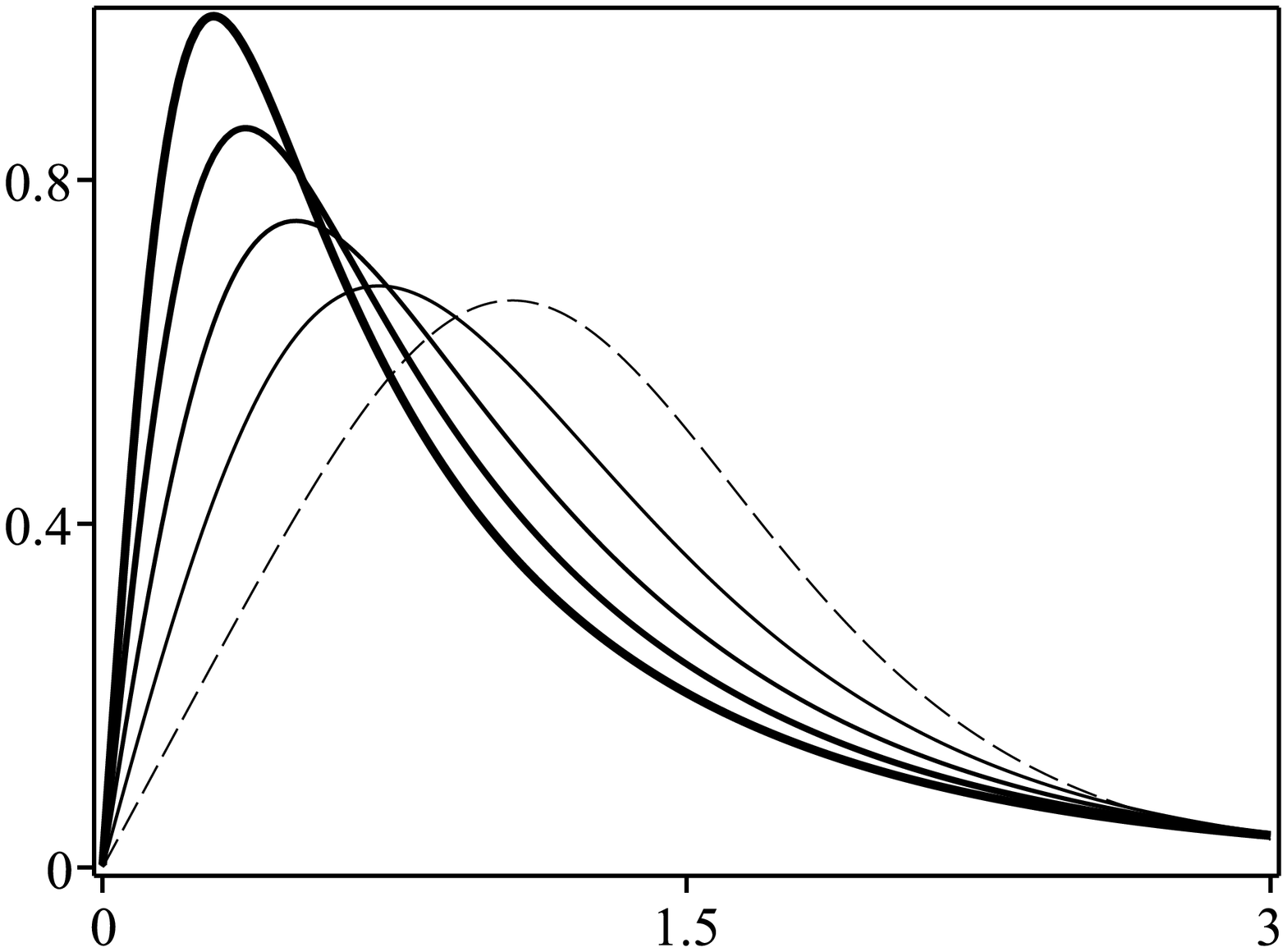}
\includegraphics[width=4.2cm,trim={0.6cm 0.2cm 0 0},clip]{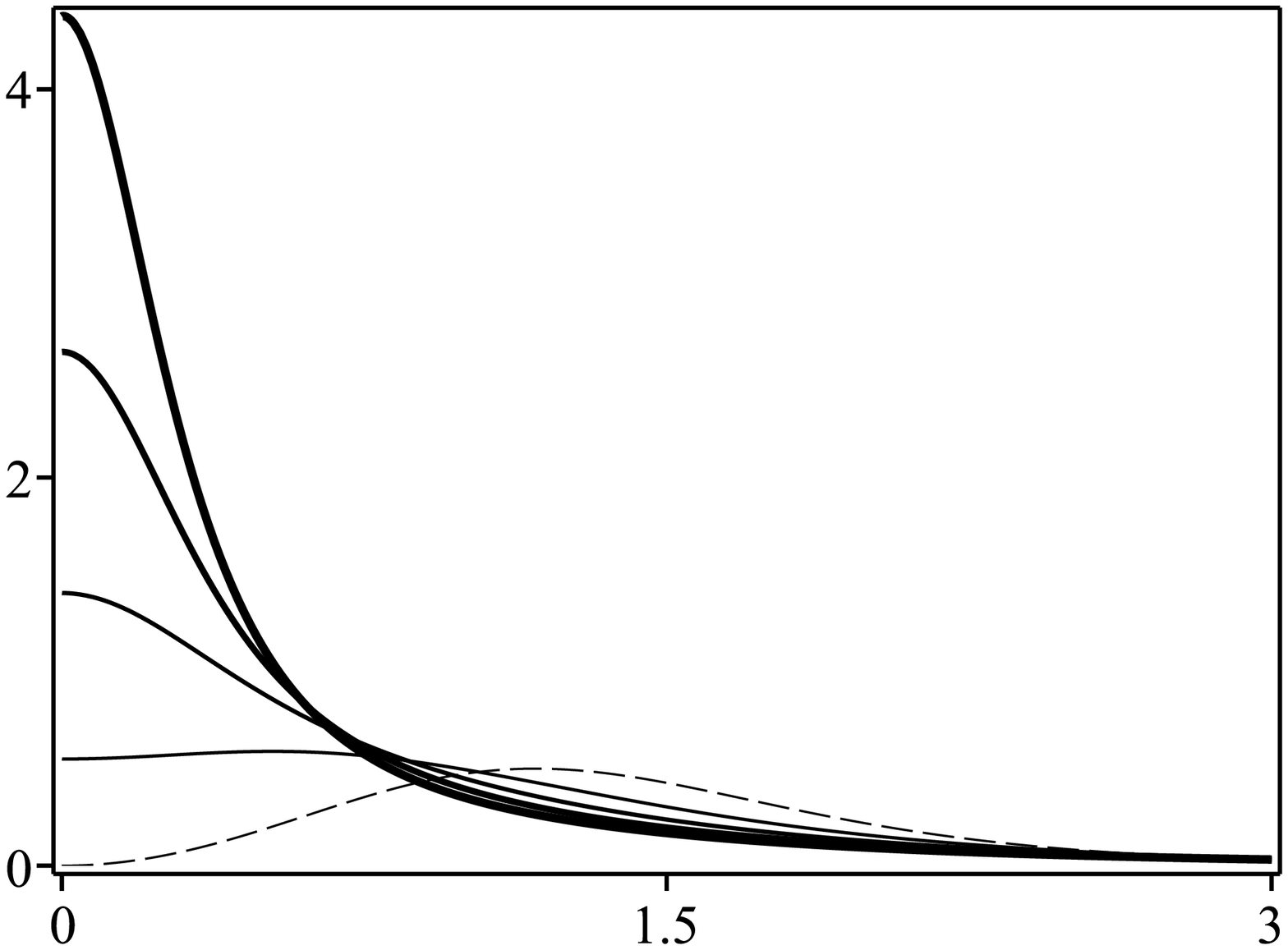}
\caption{The intensity of the electric (left) and the magnetic (right) fields in Eq.~\eqref{fields} for the solutions of Eqs.~\eqref{fo2g} and \eqref{first21a} with $e=\kappa=q=v=n=\gamma=1$, $\sigma=-1$ (top) and $1$ (bottom), and     hoiy$\lambda=1,2,3,4$. The dashed lines represent the case $\lambda=0$ and thickness of the lines increases with $\lambda$.}
\label{fig5}
\end{figure}

Unfortunately, we were not able to calculate the analytical solutions of Eqs.~\eqref{fo2g} and \eqref{first21a}. So, we use numerical procedures and display the profiles of $a(r)$, $g(r)$ and $h(r)$ for $e=\kappa=q=v=n=\gamma=1$ and some values of $\sigma$ and $\lambda$ in Fig.~\ref{fig4}. We see that, for $\sigma<0$, $a(r)$ is not monotonically decreasing: it increases near the origin until a maximum at $a_{max}>n$, and then decreases towards zero. This behavior appeared before in Ref.~\cite{burzlaff}, in a minimally coupled model with an specific modification in the Chern-Simons term. A similar behavior occurs in the function $h(r)$. The solution $g(r)$, although presents changes in the sign of its second derivative, always increases in the interval $[0,v]$. We plot the electric and magnetic fields in Fig.~\ref{fig5}. The electric field may change its sign for $\lambda>\lambda_c$, where $\lambda_c$ is as in Eq.~\eqref{lambdac}. For the magnetic field, the flip on its sign, which is an evidence of a magnetic flux inversion, occurs for any positive value of $\lambda$. This feature, although appears in the scenario of Lorentz violation in minimally coupled models (see Ref.~\cite{casanaflux}), is novel in models with nonminimal coupling. It is also of interest in condensed matter, and has appeared before in the study of fractional vortices in two-component superconductors \cite{fluxprl}. 

\begin{figure}[t!]
\centering
\includegraphics[width=4.2cm,trim={0.6cm 0.2cm 0 0},clip]{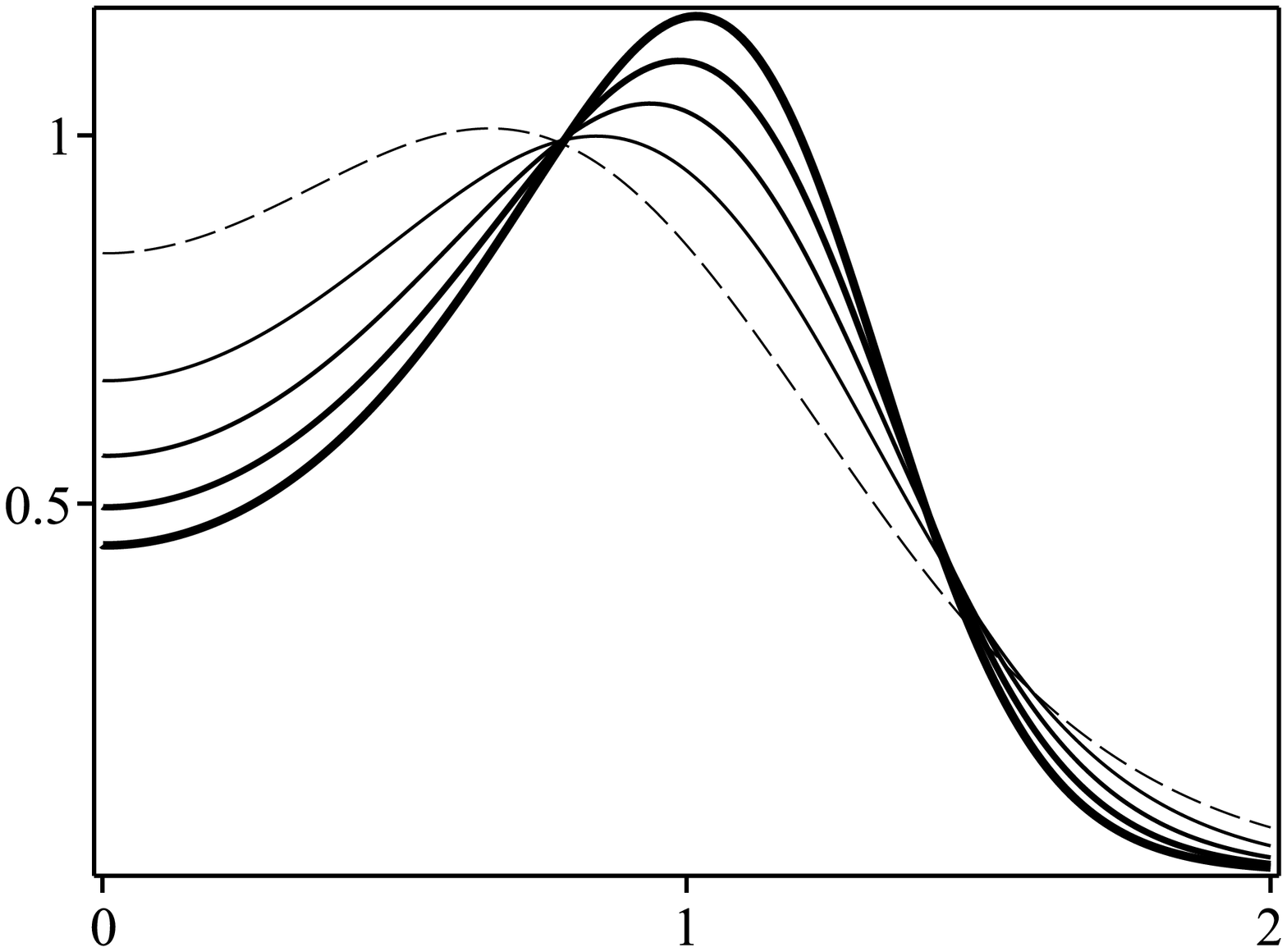}
\includegraphics[width=4.2cm,trim={0.6cm 0.2cm 0 0},clip]{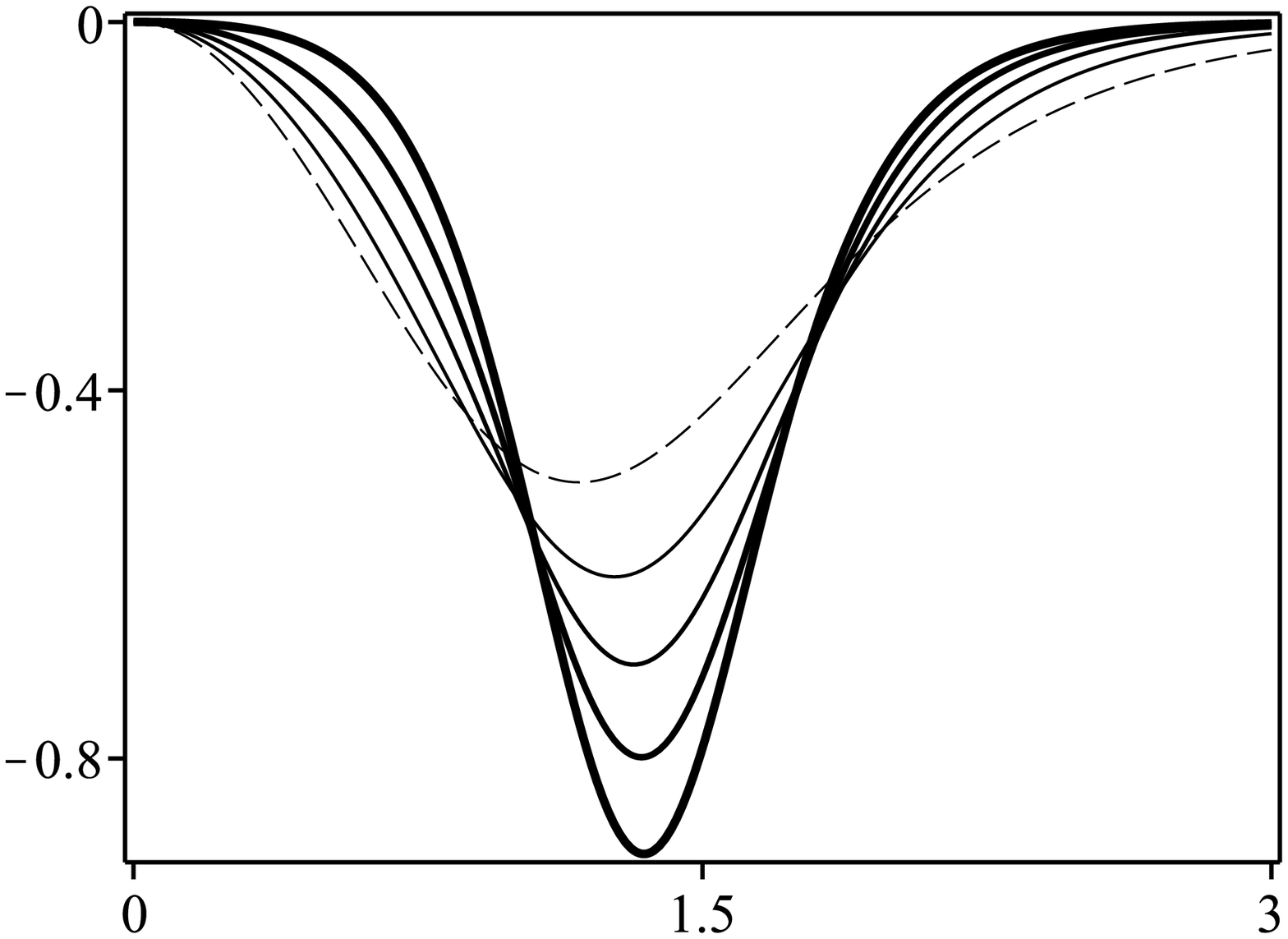}
\includegraphics[width=4.2cm,trim={0.6cm 0.2cm 0 0},clip]{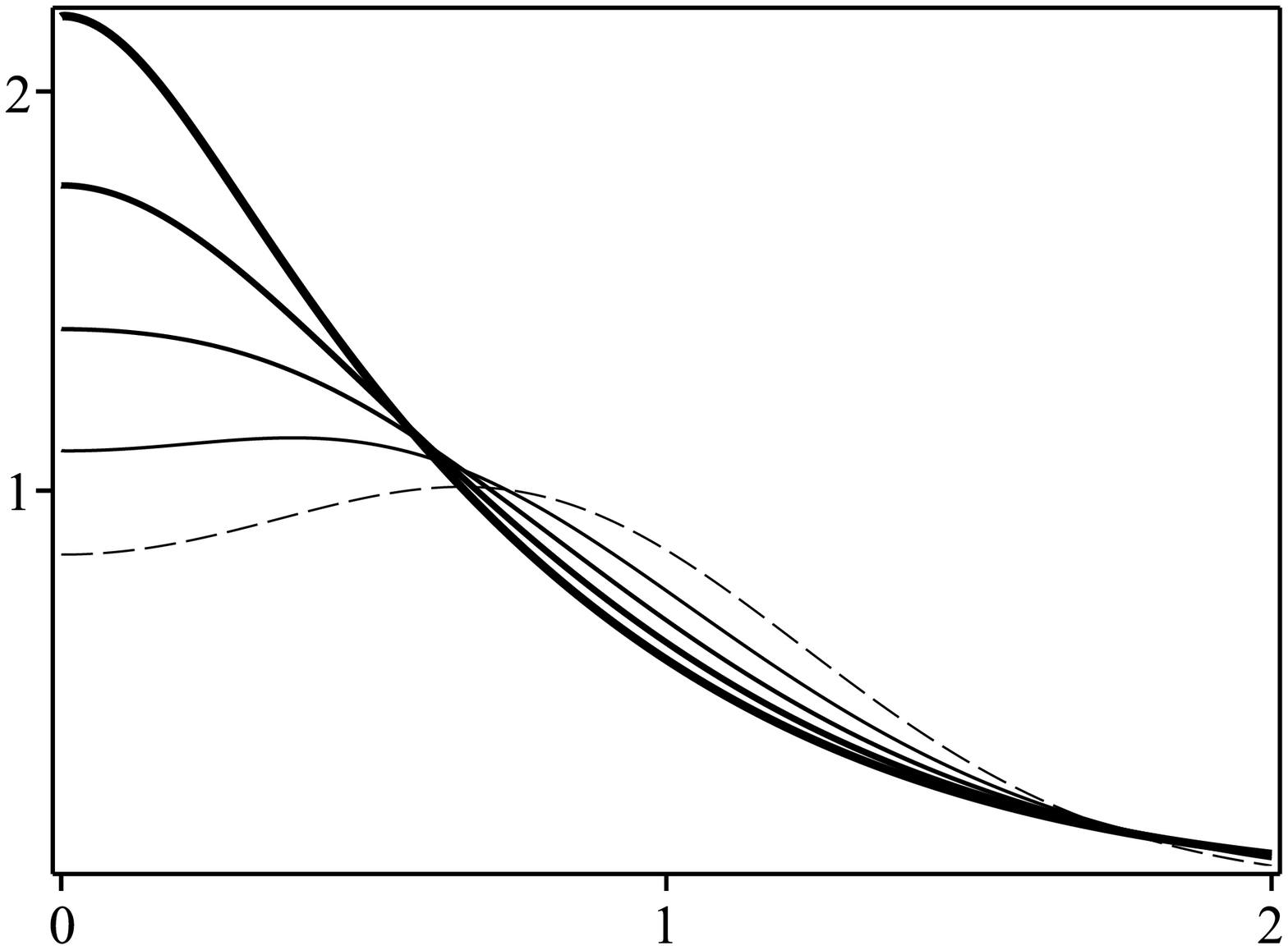}
\includegraphics[width=4.2cm,trim={0.6cm 0.2cm 0 0},clip]{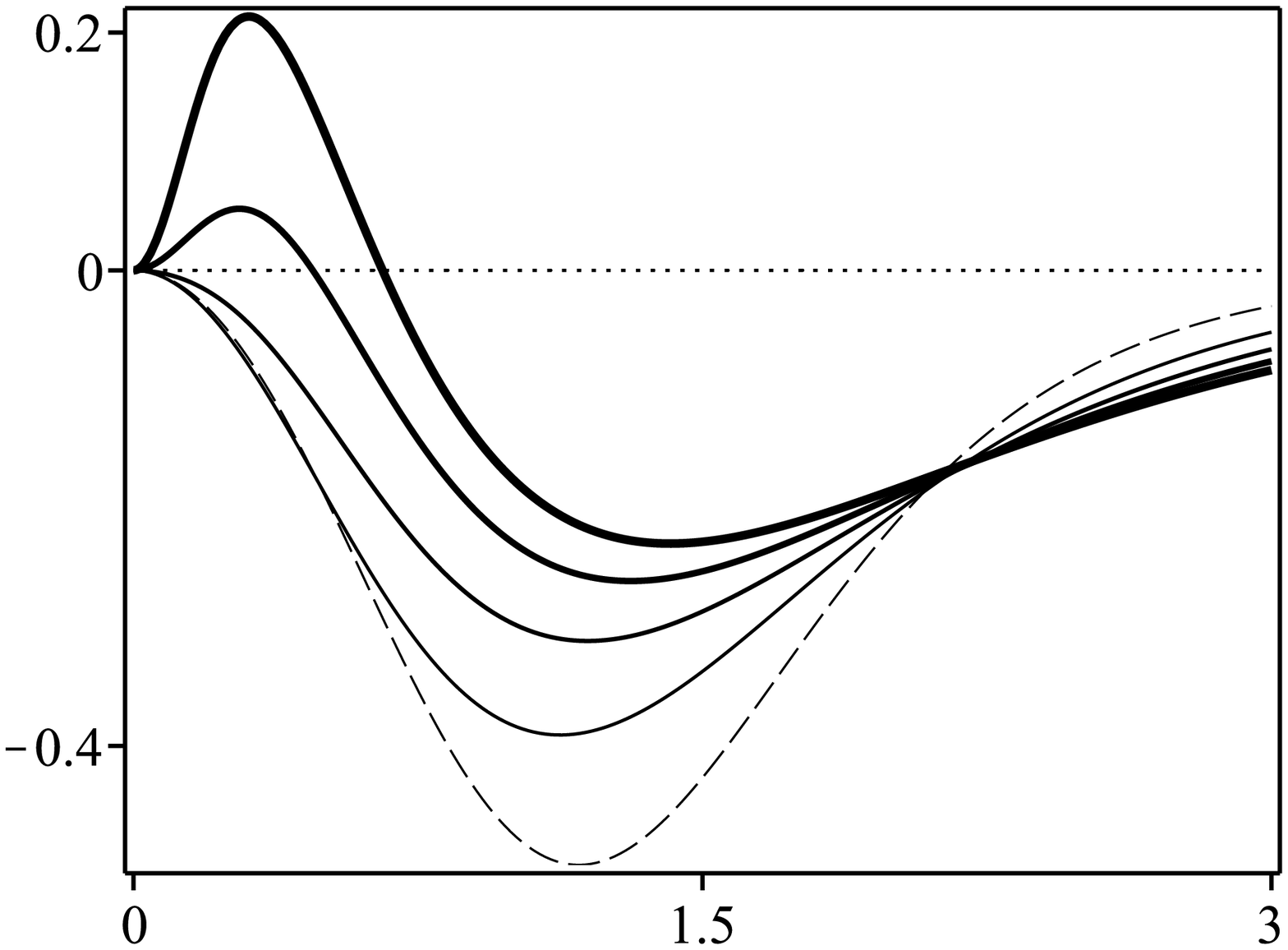}
\caption{The energy density in Eq.~\eqref{dens21} (left) and the charge density in Eq.~\eqref{j021} (right) for the solutions of Eqs.~\eqref{fo2g} and \eqref{first21a} with $e=\kappa=q=v=n=\gamma=1$, $\sigma=-1$ (top) and $1$ (bottom), $\lambda=1,2,3,4$. The dashed lines represent the case $\lambda=0$ and thickness of the lines increases with $\lambda$.}
\label{fig6}
\end{figure}

We can show that, by numerical integration for positive vorticity, the total flux is positive and quantized, $\Phi=2\pi n/e$. The energy density and the charge density can be seen in Fig.~\ref{fig6}. Notice that the energy density presents a hole around the origin that gets deeper as $\lambda$ increases. The charge density has a peak that gets taller as $\lambda$ increases. The case $\sigma>0$ leads to topological solutions with the usual monotonic behavior. However, as $\lambda$ increases, the hole in the center of both the magnetic field and energy density vanish, becoming a maximum. Moreover, the parameter $\lambda$ modifies the behavior of the charge density, which may present a change in its sign.

We now consider a distinct possibility for Eq.~\eqref{v2c}, in which $\alpha\leq0$. So, we choose $\alpha=-\beta^2$, where $\beta$ is a parameter with the dimension of square root of energy. We consider $M(\vphia)=1$, as in the models studied in Refs.~\cite{ghoshplb,ghosh}. In this situation, however, we cannot take $C=0$ as before because we would not have the proper set of minima that are connected by the solution $g(r)$ with topological nature. In this situation, to ensure the potential presents a set of minima at $g=v$, we take $C=(1 +\beta^2 v^2)^{\frac{\gamma+1}{2}}$. The potential in Eq.~\eqref{v2c} takes the form
\be
\begin{aligned}
V(g) &= \frac{4e^4g^2\left(1 +\beta^2 g^2\right)^{\gamma -1}}{\kappa^2\beta^4\left(1+\gamma\right)^2}\\
     &\times\left(\left(1 +\beta^2 v^2\right)^{\frac{\gamma+1}{2}} -\left(1 +\beta^2 g^2\right)^{\frac{\gamma+1}{2}}\right)^2,
\end{aligned}
\ee
For $\gamma=1$, we get the well-known sixth-order power law potential $V(g) = e^4g^2\left(v^2 -g^2\right)^2/\kappa^2$ that is found in the study of pure Chern-Simons models with minimal coupling \cite{jackiw1,jackiw2,coreanos}.

In the first order equations \eqref{fo2}, only Eq.~\eqref{fo2a} changes with the above potential, becoming
\be\label{first22a}
\begin{aligned}
-\frac{a^\prime}{er} &= \frac{g^2}{1 +\beta^2g^2}\Bigg(\!\!-\frac{\beta^2\!\left(\gamma-1\right)\!a^2}{er^2} +\frac{4 e^3\!\left(1 +\beta^2g^2\right)^{\frac{3\gamma-1}{2}}}{\beta^2\kappa^2\left(\gamma+1\right)}\\
	&\times\left(\!\left(1 +\beta^2 v^2\right)^{\frac{\gamma+1}{2}} -\left(1 +\beta^2 g^2\right)^{\frac{\gamma+1}{2}}\right)\!\Bigg).
\end{aligned}
\ee
By knowing the solutions $a(r)$ and $g(r)$, one may calculate the function $h(r)$ that comes from Eq.~\eqref{h2}, such that
\be\label{h22}
\begin{aligned}
h(r) &= \frac{2e\left(1 +\beta^2 g^2\right)^{\frac{\gamma-1}{2}}}{\kappa\beta^2\left(1 +\gamma\right)}\\
	&\times\left(\left(1 +\beta^2 v^2\right)^{\frac{\gamma+1}{2}} -\left(1 +\beta^2 g^2\right)^{\frac{\gamma+1}{2}}\right).
\end{aligned}
\ee
So, we have $h(0) = 2e\left(1 +\beta^2 v^2\right)^{\frac{\gamma+1}{2}}\!\!\big/\!\left(\kappa\beta^2(1 +\gamma)\right)$. As in the previous example, we check if the function $h(r)$ supports a global maximum at some point. By taking $h_g=0$, we get
\be
\tilde{g} = \frac{1}{\beta} \sqrt{(1+\beta^2v^2) \left(\frac{\gamma-1}{2\gamma}\right)^{\frac{2}{\gamma+1}}-1}.
\ee
So, for values of $\beta$ and $\gamma$ that lead to $0<\tilde{g}<v$, the function $h(r)$ present null derivative, defining a point of maximum. This condition is attained for $\beta>\beta_c$, with
\be\label{betamin}
\beta_c = \frac1v\sqrt{\left(\frac{2\gamma}{\gamma-1}\right)^{\frac{2}{\gamma+1}}-1}.
\ee
For $\beta$ in the aforementioned range, the electric field engender a change of sign due to the existence of a maximum in $h$.

The auxiliary function $W(a,g)$ in Eq.~\eqref{wtrick2} takes the form
\be\label{w22}
\begin{aligned}
W(a,g) &= -\frac{2a\left(1 +\beta^2g^2\right)^{\frac{1 -\gamma}{2}}}{\beta^2\left(1 +\gamma\right)}\\
	&\times\left(\left(1 +\beta^2 v^2\right)^{\frac{\gamma+1}{2}} -\left(1 +\beta^2 g^2\right)^{\frac{\gamma+1}{2}}\right).
\end{aligned}
\ee
One can combine Eqs.~\eqref{rho2}, \eqref{fo2g} and \eqref{first22a} to show that the energy density can be written as
\be\label{dens22}
\begin{aligned}
\rho &= 2g^2\Bigg(\frac{a^2}{r^2} +\frac{4e^2\left(1 +\beta^2g^2\right)^{\gamma -1}}{\kappa^2\beta^4\left(1 +\gamma\right)^2}\\
	&\times\left(\left(1 +\beta^2 v^2\right)^{\frac{\gamma+1}{2}} -\left(1 +\beta^2 g^2\right)^{\frac{\gamma+1}{2}}\right)^2\Bigg).
\end{aligned}
\ee
We can see the above energy density leads to finite energy for both topological and nontopological solutions. As before, we only deal with topological configurations, which require $a(\infty) \to 0$ and $g(\infty)\to v$ to attain the finite character of the energy. These boundary conditions may be used with Eqs.~\eqref{w22} and \eqref{ebtrick1} to show that the topological solutions have energy $E=4\pi |n| ((1+\beta^2v^2)^{\frac{\gamma+1}{2}}-1)/(\beta^2(\gamma+1))$. Differently from the previous example, one can integrate the magnetic field in Eq.~\eqref{fields} to show the magnetic flux is quantised regardless the values of the parameter $\gamma$, such that $\Phi=2\pi n/e$. The charge density in Eq.~\eqref{j02} takes the form
\be\label{j022}
J_0 = \frac{\kappa}{2er}\left(a+a\left(1+\beta^2\,g^2\right)^{1-\gamma}\right)^\prime.
\ee
By integrating the above expression, one can show the topological solutions engender charge $Q=-2\pi\kappa n/e$.

\begin{figure}[t!]
\centering
\includegraphics[width=4.2cm,trim={0.6cm 0.2cm 0 0},clip]{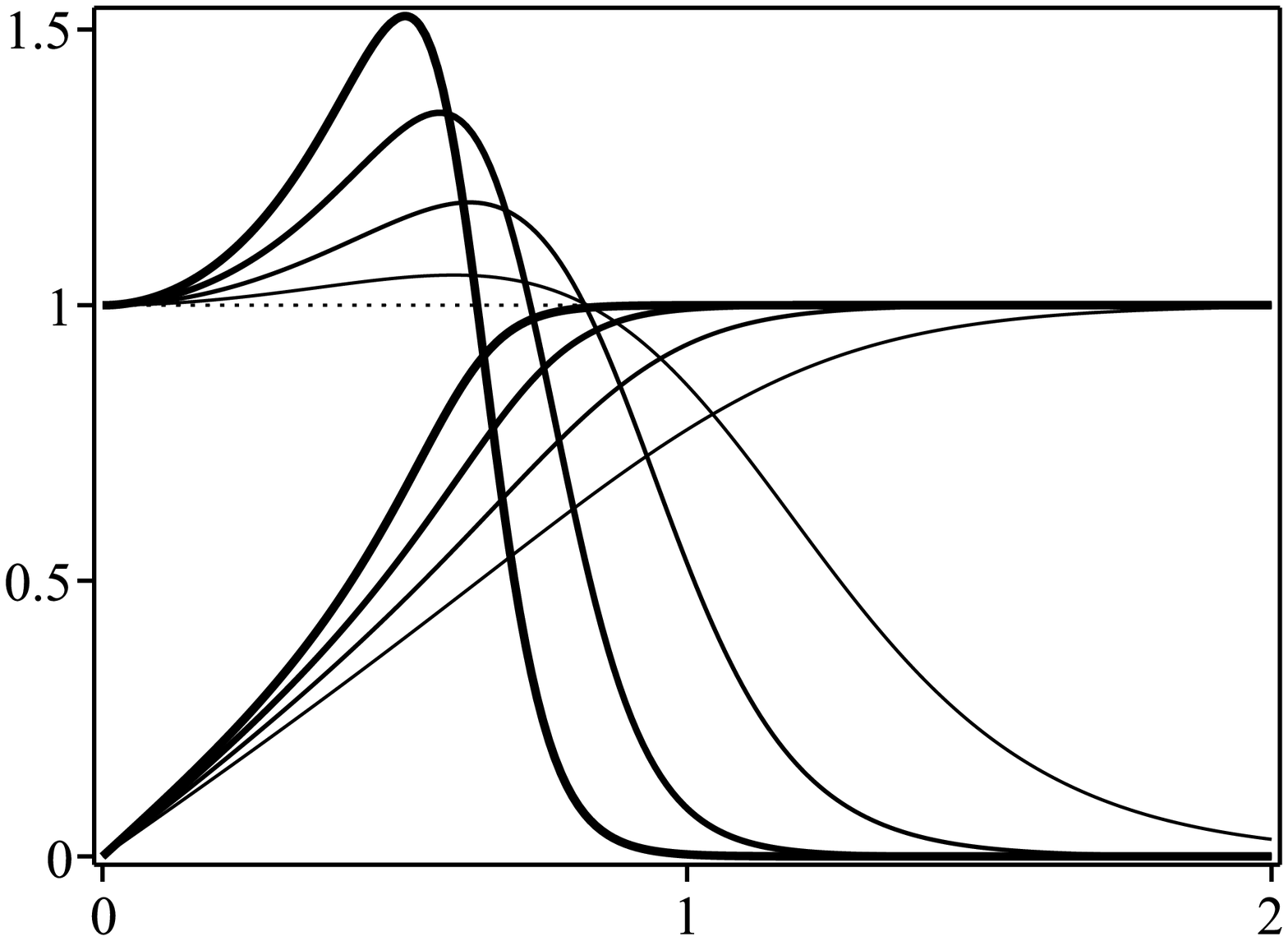}
\includegraphics[width=4.2cm,trim={0.6cm 0.2cm 0 0},clip]{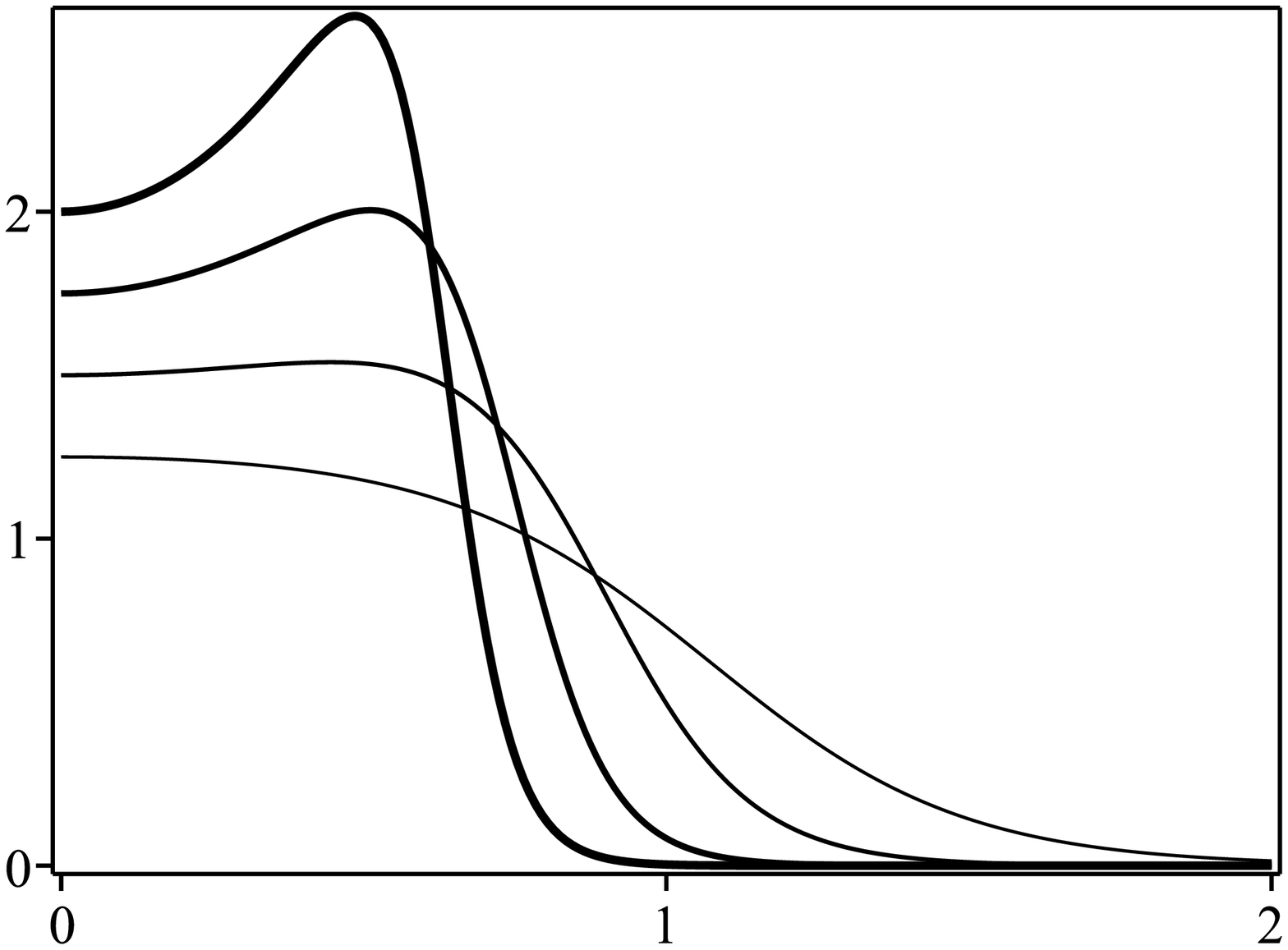}
\caption{The solutions $a(r)$ and $g(r)$ of Eqs.~\eqref{fo2g} and \eqref{first22a} (left) and the function $h(r)$ in Eq.~\eqref{h22} (right) for $e=\kappa=q=v=n=1$, $\gamma=3$ and $\beta^2=0.5,1, 1.5$ and $2$. The thickness of the lines increases with $\beta$.}
\label{fig7}
\end{figure}

We were not able to find the analytical solutions of the first order equations \eqref{fo2g} and \eqref{first22a}. So, we use numerical methods and plot the profiles of $a(r)$, $g(r)$ and $h(r)$ in Fig.~\ref{fig7} for $e=\kappa=q=v=n=1$, $\gamma=3$ and some values of $\beta$. Notice that, similarly to the previous model, we see that $a(r)$ is not monotonically decreasing as usual. Near the origin, it increases as $r$ gets larger, until it reaches a maximum value and then starts decreasing towards the boundary condition. This behavior becomes more evident as one increases $\beta$. As we have commented before, the behavior of $g(r)$ for configurations with $n=1$ near the origin is $g(r)\propto r$ (see Eq.~\eqref{gori}). However, we see that, as $\beta$ increases, the change in the second derivative along its path becomes more visible, presenting an inflection point. Regarding the function $h(r)$, as we have explained above, it has a minimum at $r\neq0$ for $\beta>\beta_c$, with $\beta_c$ as in Eq.~\eqref{betamin}. 

\begin{figure}[t!]
\centering
\includegraphics[width=4.2cm,trim={0.6cm 0.2cm 0 0},clip]{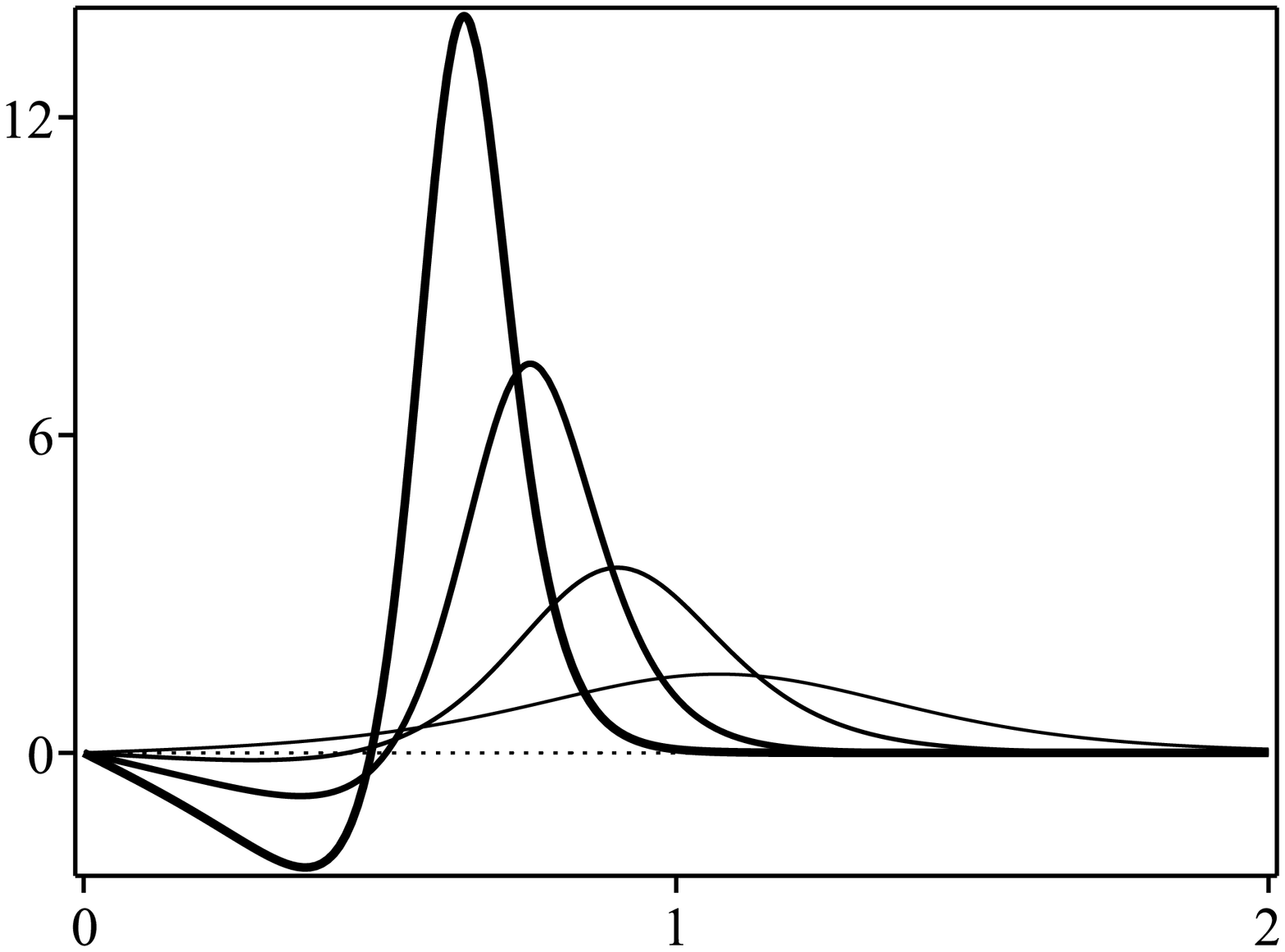}
\includegraphics[width=4.2cm,trim={0.6cm 0.2cm 0 0},clip]{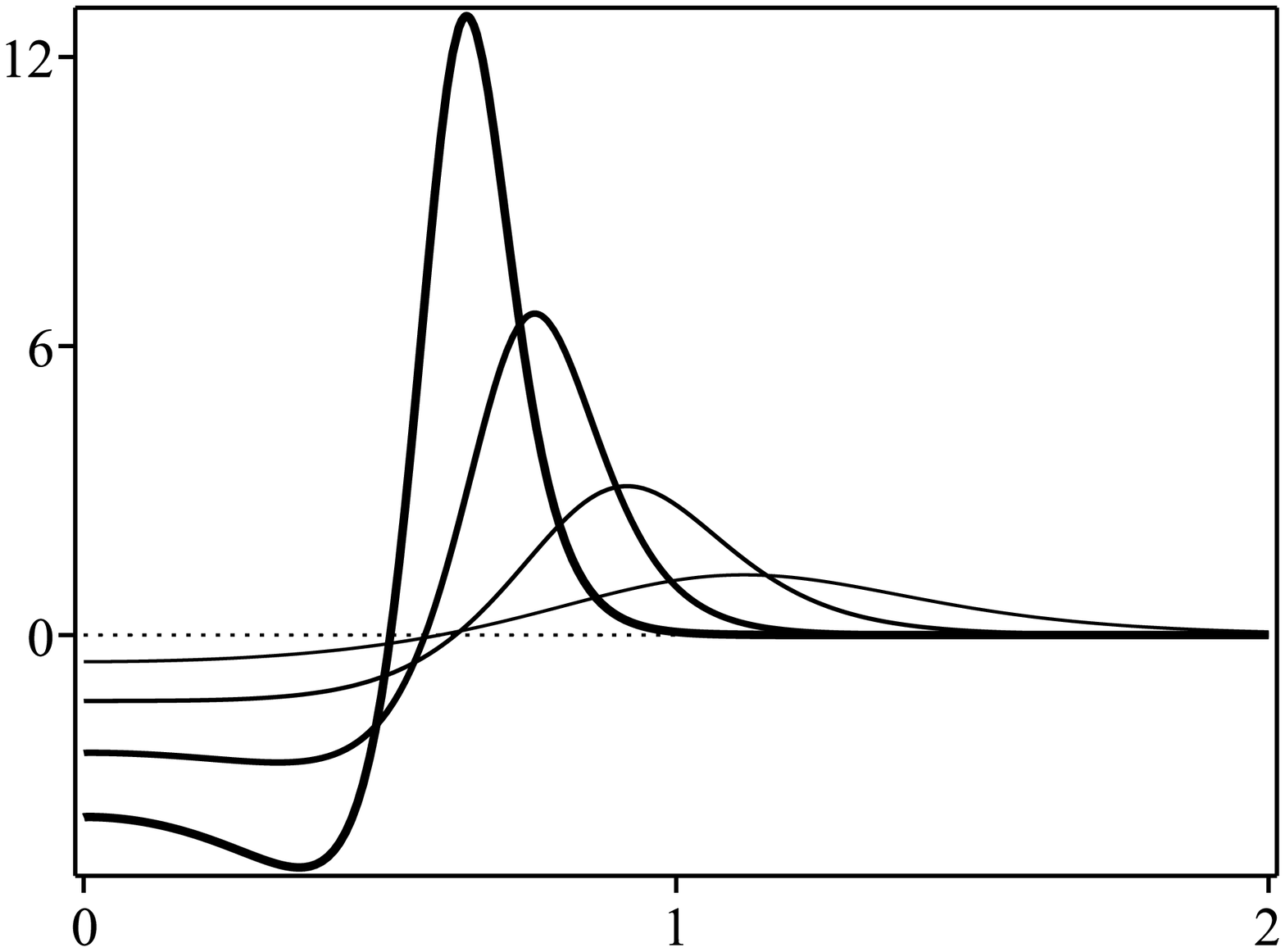}
\caption{The intensity of the electric (left) and the magnetic (right) fields in Eq.~\eqref{fields} for the solutions of Eqs.~\eqref{fo2g} and \eqref{first22a} with $e=\kappa=q=v=n=1$, $\gamma=3$ and $\beta^2=0.5,1, 1.5$ and $2$. The thickness of the lines increases with $\beta$.}
\label{fig8}
\end{figure}

\begin{figure}[t!]
\centering
\includegraphics[width=4.2cm,trim={0.6cm 0.2cm 0 0},clip]{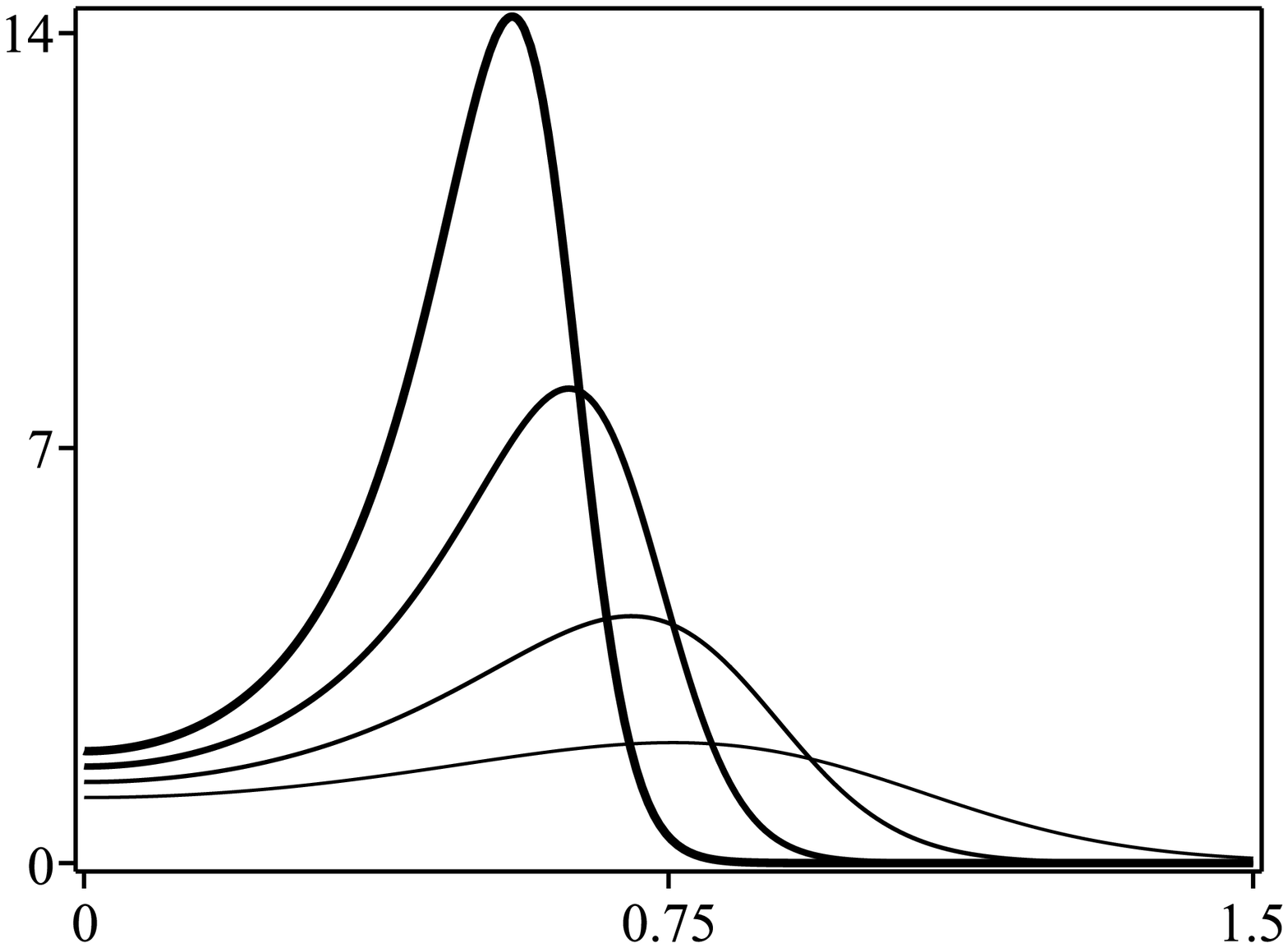}
\includegraphics[width=4.2cm,trim={0.6cm 0.2cm 0 0},clip]{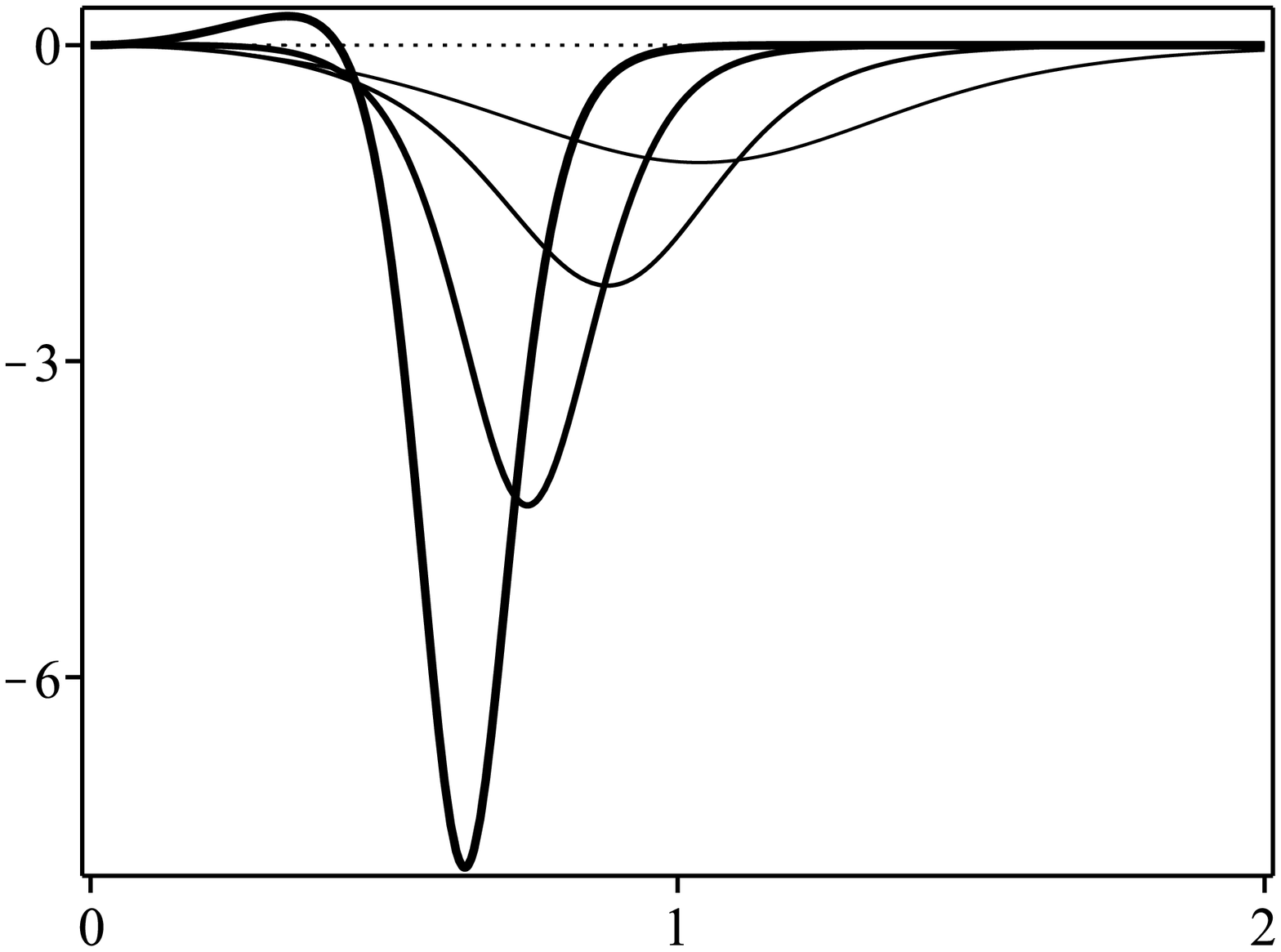}
\caption{The energy density in Eq.~\eqref{dens22} (left) and the charge density in Eq.~\eqref{j022} (right) for the solutions of Eqs.~\eqref{fo2g} and \eqref{first22a} with $e=\kappa=q=v=n=1$, $\gamma=3$ and $\beta^2=0.5,1, 1.5$ and $2$. The thickness of the lines increases with $\beta$.}
\label{fig9}
\end{figure}

By making use of the aforementioned solutions, we also plot the corresponding electric and magnetic fields \eqref{fields} in Fig.~\ref{fig8}. Notice the magnetic field is negative around the origin, with a valley getting deeper as $\beta$ increases. This means that the vortex engender a magnetic flux inversion. Notwithstanding that, the total flux is positive and quantized, given by $\Phi = 2\pi n/e$. In the electric field, the inversion of sign occurs only for $\beta>\beta_c$, with $\beta_c$ as in Eq.~\eqref{betamin}. The energy density \eqref{dens22} and the charge density \eqref{j022} are plotted in Fig.~\ref{fig9}. Notice that the charge density may also engender a change of sign, depending on the $\beta$ chosen, whilst the energy density is always non negative.

\section{Conclusion}\label{end}

In this paper, we have investigated vortex configurations in a class of generalized Maxwell-Chern-Simons models with a complex scalar field nonminimally coupled to the gauge field. The general model is described by the Lagrange density \eqref{lagrange}, which, in addition to the potential $V(\vphia)$, presents the functions $P(\vphia)$ that controls a generalized magnetic permeability, $M(\vphia)$ which was not introduced in previous works and drives the dynamical term of the scalar field, and $G(\vphia)$, that controls the term that gives rise to the nonminimal coupling. The main properties are calculated, such as the equations of motion, the current and the energy-momentum tensor. By considering static configurations with the fields given as in Eq.~\eqref{ansatz}, we show the equations of motion are of second order. In order to simplify the problem, we focused on developing a first order formalism to describe the configurations of interest.

First, we have followed the suggestion described in Refs.~\cite{torres,ghoshplb,ghosh}, considering the condition $J^\mu = \kappa F^\mu$, which imposes a constraint between $P(\vphia)$ and $G(\vphia)$. In this situation, we showed the electric charge is related to the magnetic flux. Then, we developed the Bogomol'nyi procedure for the model, which allowed us to find first order equations whose solutions minimizes the energy of the system and are compatible with the equations of motion. By taking specific functions $G(\vphia)$ and $M(\vphia)$, we introduced a novel model that modifies the behavior of the $g(r)$ near the origin, which engenders a plateau whose width is controlled by a parameter in the function $M(\vphia)$. Moreover, oppositely to the model in Ref.~\cite{ghoshplb,ghosh}, the magnetic permeability is non negative. This model support magnetic and electric fields, and the energy density with a ringlike shape whose internal radius is governed by the aforementioned parameter.

We have also introduced a novel manner to obtain a first order formalism, in which $P(\vphia)$ is constrained by $G(\vphia)$ and $M(\vphia)$. In this case, the Bogomol'nyi procedure is also developed. So, we get minimal energy configurations that comes from first order equations compatible with the equations of motion. Interestingly, the first order equation \eqref{fo2g} is the very same of the one that arises in the study of vortices in models with minimal coupling, such as the ones in Refs.~\cite{NO,bogopaper,jackiw1,jackiw2,coreanos,godvortex}. So, near the origin, there is only one possible behavior for $g$, in the form $g(r\approx0)\propto r^{|n|}$. On the other hand, the first order equation \eqref{fo2a} brings a novel feature to the problem: the presence of a term with $a(r)$. This new term competes with the one that depends on the potential and may causes significant changes in the profile of $a(r)$. We then provided specific examples in which $a(r)$ is not monotonically decreasing and both the magnetic and electric fields may present a change of sign. Even though a magnetic flux inversion occurs, the total flux is positive. Moreover, these unusual features does not modify the positiveness of the energy density, such that the energy is positive.

There are several distinct possibilities of extending the present work, among them the case of vortices controlled by non Abelian gauge symmetries \cite{V}, the presence of magnetic monopoles with Abelian charges \cite{W}, the case of nonrelativistic dynamics \cite{nm4,nm10,JP,Hor} and the study of vortices in Bose-Einstein condensates \cite{BE1,BE2}. 

\acknowledgements{The work is supported by the Brazilian agencies Coordena\c{c}\~ao de Aperfei\c{c}oamento de Pessoal de N\'ivel Superior (CAPES), grant No.~88887.463746/2019-00 (MAM), Conselho Nacional de Desenvolvimento Cient\'ifico e Tecnol\'ogico (CNPq), grants Nos. 140490/2018-3 (IA), 303469/2019-6 (DB), 404913/2018-0 (DB) and 306504/2018-9 (RM), and by Paraiba State Research Foundation (FAPESQ-PB) grants Nos. 0003/2019 (RM) and 0015/2019 (DB).}

\end{document}